\documentclass[aps,prb,twocolumn,showpacs,groupedaddress]{revtex4}
\usepackage{graphicx}
\usepackage{amssymb}
\usepackage{amsmath}
\usepackage{xcolor}
\newlength{\figwidth}
\newcommand{\ud}{\mathrm{d}}
\usepackage{epsf}
\begin{document}
\preprint{\today}
\title{Scanning Gate Microscopy of Quantum Contacts Under Parallel Magnetic Field:
\\ Beating Patterns Between Spin-Split Transmission Peaks or Channel Openings}
 
\author{Andrii Kleshchonok
\footnote{Present Address: Institute for Materials Science, Dresden University of Technology, Hallwachstr. 3, 01062 Dresden, Germany; \\
Permanent Address: Taras Shevchenko National University of Kyiv, Physics Faculty, 64, Vladymirska St., 01601 Kyiv, Ukraine}}
\affiliation{Service de Physique de l'\'Etat Condens\'e (CNRS UMR 3680), IRAMIS/SPEC, CEA Saclay, 91191 Gif-sur-Yvette, France}

\author{Genevi\`eve Fleury}
\affiliation{Service de Physique de l'\'Etat Condens\'e (CNRS UMR 3680), 
 IRAMIS/SPEC, CEA Saclay, 91191 Gif-sur-Yvette, France}

\author{Gabriel Lemari\'e}
 \affiliation{Laboratoire de Physique Th\'eorique, UMR-5152, CNRS and Universit\'e de Toulouse, F 31062 France}

\author{Jean-Louis Pichard}
\affiliation{Service de Physique de l'\'Etat Condens\'e (CNRS UMR 3680), IRAMIS/SPEC, CEA Saclay, 91191 Gif-sur-Yvette, France}

\begin{abstract} 
 We study the conductance $g$ of an electron interferometer created in a two dimensional electron gas 
between a nanostructured contact and the depletion region induced by the charged tip of a scanning gate microscope. 
Using non-interacting models, we study the beating pattern of interference fringes exhibited by the images giving $g$ 
as a function of the tip position when a parallel magnetic field is applied. The analytical solution of a simplified 
model allows us to distinguish between two cases: (i) If the field is applied everywhere, the beating of Fabry-P\'erot 
oscillations of opposite spins gives rise to interference rings which can be observed at low temperatures when the 
contact is open between spin-split transmission resonances. (ii) If the field acts only upon the contact, the interference 
rings cannot be observed at low temperatures, but only at temperatures of the order of the Zeeman energy. For a 
contact made of two sites in series, a model often used for describing an inversion-symmetric double-dot setup, 
a pseudo-spin degeneracy is broken by the inter-dot coupling and a similar beating effect can be observed without magnetic 
field at temperatures of the order of the interdot coupling. Eventually, numerical studies of a quantum point contact 
with quantized conductance plateaus confirm that a parallel magnetic field applied everywhere or only upon the contact 
gives rises to similar beating effects between spin-split channel openings. 
\end{abstract}   
\pacs{
      07.79.-v, 
72.10.-d  
73.63.Rt  
}
 
\maketitle

\begin{figure}
\includegraphics[keepaspectratio,width=\columnwidth]{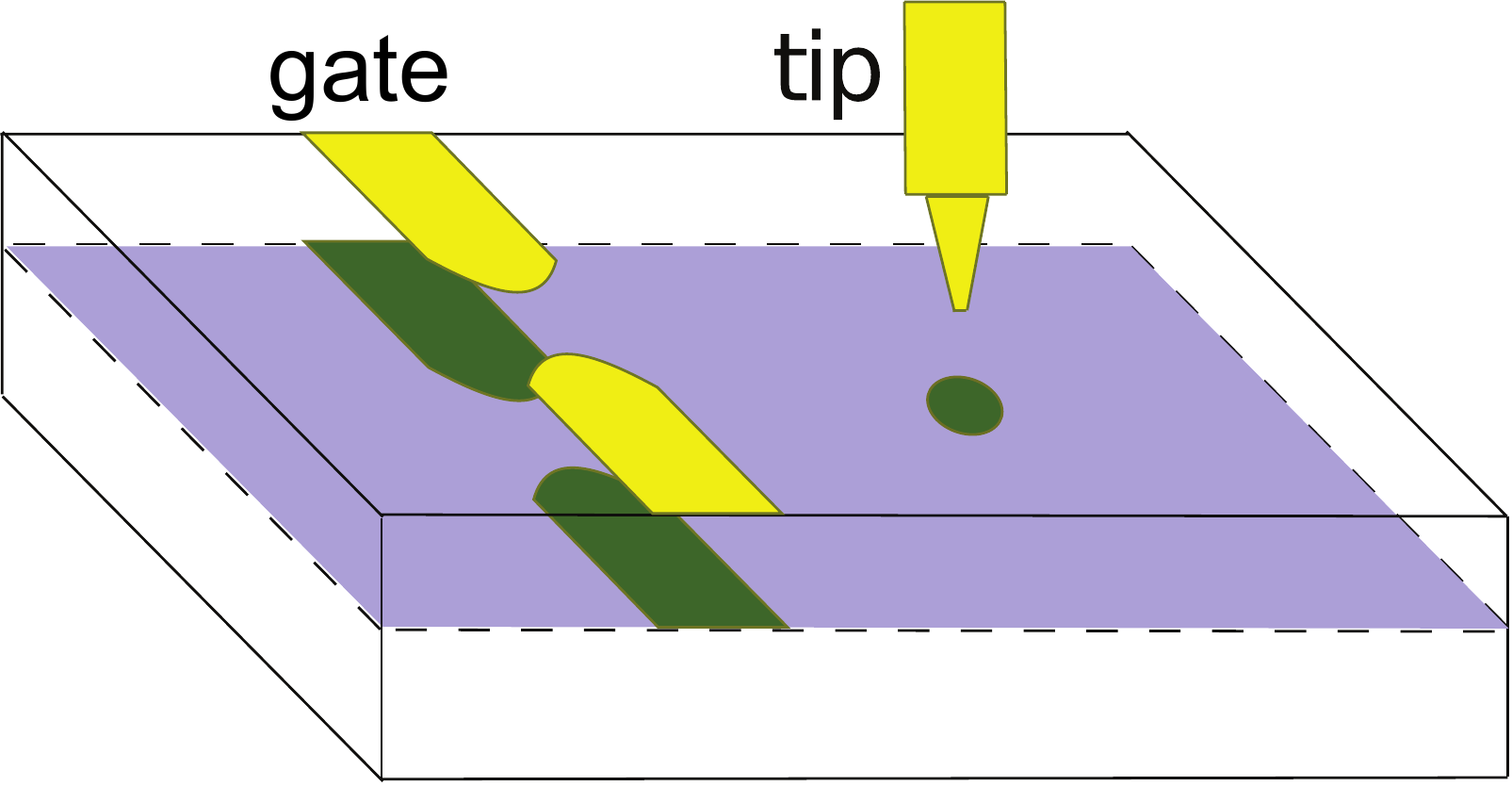}
\caption{(Color online) Scanning gate microscopy of a quantum contact: Metallic gates (yellow) create 
in the 2DEG (blue) beneath the surface a nanostuctured contact (green) which divides the 2DEG into a left 
and right parts. The depletion region induced by a scannable charged tip (green disk) and the contact form 
a 2D electron interferometer. The SGM images give the interferometer conductance as a function of the tip 
position. 
} 
\label{fig1}
\end{figure}

 Scanning gate microscopy (SGM) is a tool for probing by electron interferometry~\cite{Topinka:PT03} the properties 
of nanostructures created in a two-dimensional electron gas (2DEG). Using charged gates deposited on the surface of a semi-conductor 
heterostructure, one can divide the 2DEG beneath the surface in two parts connected via a more or less simple contact region.
This region can be a quantum point contact~\cite{PhysRevLett.60.848,PhysRevLett.77.135} (QPC) as sketched in 
Fig.~\ref{fig1}, a quantum dot~\cite{nature391,science281,Goldhaber-gordon}, a double-dot setup~\cite{Molenkamp:PRL95,revmodphys} 
or more complex nanostructures. With the charged tip of an atomic force microscope free to move above the surface of the heterostructure, 
a depletion region can be capacitively induced in the 2DEG below the surface at a distance $r$ from the contact. By scanning the tip 
outside the contact, one can record SGM images where a color code gives the conductance $g$ of the resulting electron interferometer 
as a function of the tip position. These images exhibit Fabry-P\'erot interference fringes spaced by half the Fermi wavelength $\lambda_F/2$, 
as first observed by Topinka et al~\cite{Topinka:Sci00} using a QPC opened on its first conductance plateau. This has led to revisit the 
theory of the 2D electron interferometers~\cite{Heller:NL05,metalidis,fkp,Jalabert:PRL10,alp,gorini1,gorini2,nowak} made with a QPC. 
These studies have shown a rich variety of interference phenomena, which depend on the opening of the contact, on the 
presence of electron-electron interaction effects~\cite{fkp} inside the contact, and which can exhibit a non trivial temperature dependence, 
as pointed out in Ref.~\cite{alp}. At the same time, experimental SGM studies of QPCs performed at lower temperatures have revealed unexpected 
behaviors: (i) The interference fringes at $300 mK$ can exhibit~\cite{ensslin} enhanced and reduced contrasts as the distance $r$ between the QPC 
and the tip increases; (ii) A recent SGM study~\cite{brun} at $20 mK$ shows that one can control the $0.7 \times 2e^2/h$ anomaly of a QPC with a 
scanning gate, revealing possible relations with Wigner and Kondo physics. 
In this paper, we discuss novel quantum interference effects which characterize non interacting models of interferometers describing the 
SGM of quantum contacts when a parallel magnetic field is applied (the role of electron-electron interaction 
inside the contact can be important~\cite{fkp,brun} and will be considered in a second paper). To introduce these effects, we use a 
simplified lattice model (see section I), the resonant contact model (RCM) which was previously used in Ref.~\cite{alp}. In the RCM model, 
the contact region between two semi-infinite square lattices and the depletion region induced by the charged tip are reduced to single lattice sites 
(see fig.~\ref{fig2QPC}). This makes the RCM model analytically solvable (see subsections II.A and II.B). Its transmission without tip exhibits a 
single spin-degenerate Breit-Wigner resonance as one varies the energy $E$.\\ 
\indent In Ref.~\cite{alp}, we have studied thermally induced interference fringes which characterize the SGM of such a contact when it is opened around 
its spin-degenerate resonance. Here, we study other interference effects occuring when the spin degeneracy of the transmission resonance is removed 
by a parallel magnetic field. Plotting the SGM images when the contact is open between the two spin-split resonances, one expects to detect a beating 
effect between the two spin contributions to the interferometer conductance. If the magnetic field is applied everywhere (contact and 2DEG), the 
pattern of interference rings induced by the field can be mainly observed as the temperature ${\cal T} \to 0$, the radii of the rings being independent 
of ${\cal T}$ (see subsection II.C).\\
\indent If the field is applied upon the contact only (and not upon the 2DEG) a non-trivial effect is given by the analytical solution of 
the RCM model: the observation of the rings requires a temperature of the order of the Zeeman splitting. This is due to the temperature dependence 
of their radii. The rings cannot be seen when ${\cal T} \to 0$ because their radii become infinite. This is only when one increases 
${\cal T}$ that the radii become sufficiently small and that the beating effect can be seen in the SGM images (see subsection II. D).\\ 
\indent This thermally induced interference effect can also be seen without magnetic field, in a contact exhibiting a double-peak structure of 
its spin-degenerate transmission function. We study in section III such an example where the contact is made of two sites in series instead of a 
single one. This model is suitable for modeling a contact made of a double-dot setup and exhibits two transmission peaks without magnetic field. 
When the two dots are identical and decoupled, there is nevertheless a pseudo-spin degeneracy caused by the inversion symmetry of the model, and 
the inter-site hopping term of the double-dot setup plays the role of the Zeeman energy. The SGM images exhibit also a thermally induced beating 
pattern of interference fringes at temperatures of the order of the inter-site hopping term.
\\ 
\indent In those two examples, the beating  effect between Fabry-P\'erot interference fringes of opposite spins or pseudo-spins gives rise to a pattern 
of rings with a characteristic scale much larger than $\lambda_F/2$. This comes from the difference of phases (and not of amplitudes) induced by 
the Zeeman energy between the two spin contributions to $g$. These thermally induced beating phenomena should be distinguished from the more trivial 
beating effect induced by a global parallel magnetic field acting upon the whole interferometer. In this last case, the amplitudes (and the phases) 
of the two spin contributions to $g$ are different and the rings remain visible when ${\cal T} \to 0$. Measuring the spacing between the rings, one 
can extract either the Zeeman energy for the single site contact, or the inter-site coupling for the double-site contact.\\ 
\indent We eventually show in section IV that the thermally induced interference phenomena which can be analytically described using the RCM model remain 
relevant for saddle-point contacts having a staircase energy dependence of their transmission with quantized number of transmission channels. This 
is numerically illustrated taking a QPC opened near a channel opening, when the spin degeneracy is removed by a parallel magnetic field. As in 
the RCM model, a local field inside the QPC gives rise to a different temperature dependence of the beating pattern than if the field is applied 
everywhere. 

\section{Lattice Models for Quantum Contacts}
\begin{figure}
\includegraphics[keepaspectratio,width=\columnwidth]{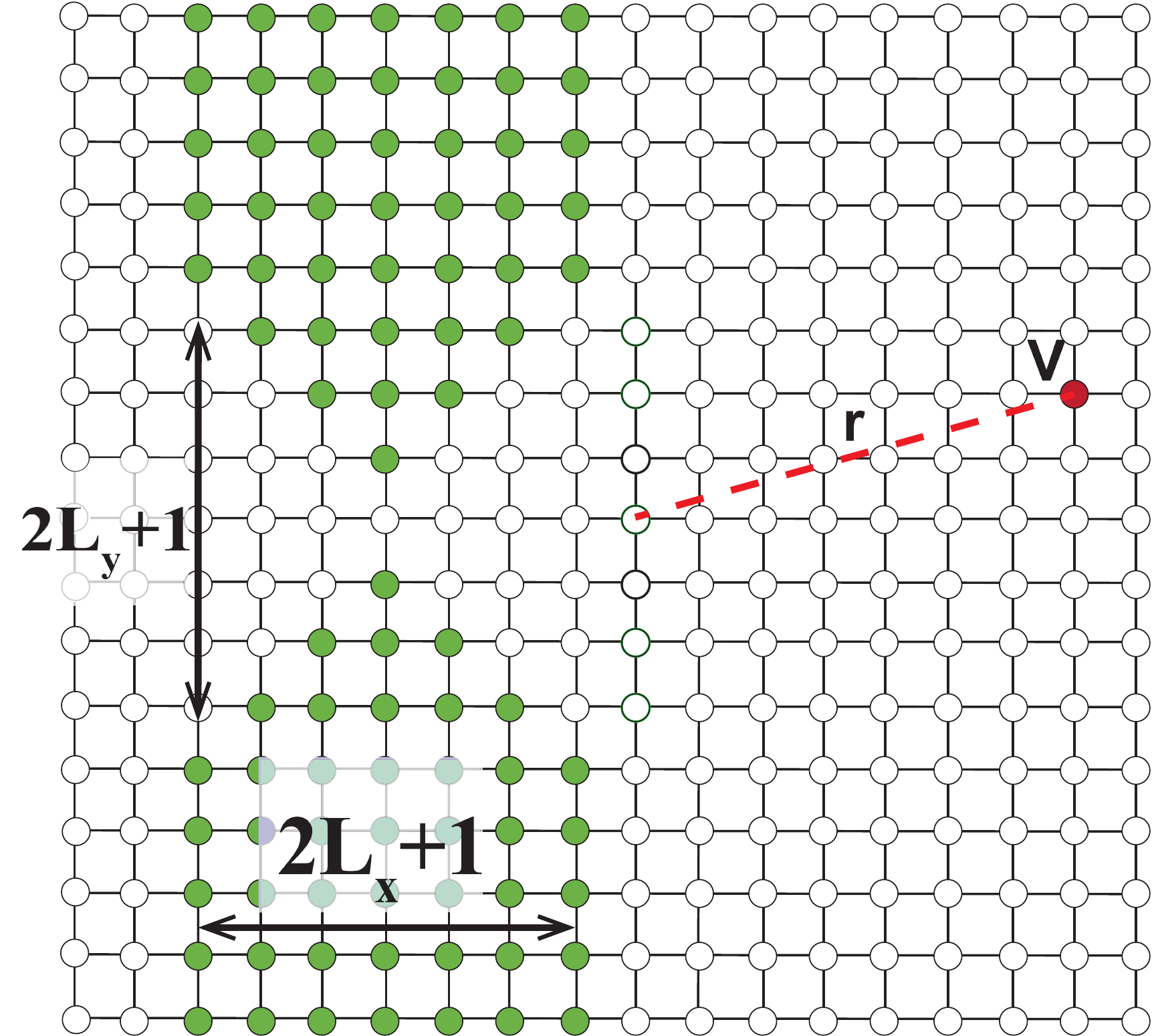}
 \caption{(Color online) Lattice model for the SGM of a QPC: Two semi-infinite square lattices (leads) are contacted by a region of length 
$2L_x+1$ and width $2L_y+1$. The site potentials are equal to $-4t$, excepted in the contact region ($-L_x\leq i_x \leq L_x$) where the 
potential at a site ${\bf i}$ of coordinates $(i_x,i_y)$ is taken infinite (green sites) if $|i_y| \geq (L_y-k)+k \left(i_x/L_x \right)^2$ 
($L_x=L_y=3$ and $k=2$) and in a single site (red) of coordinates $(x,y)$ and of potential $V-4t$ which describes the depletion region 
induced by the charged tip.}
\label{fig2QPC}
\end{figure}

For studying the 2D interferometer formed in a 2DEG between a QPC and the depletion region induced by a charged tip,  
we use lattice models where two semi-infinite square lattices (leads) are connected by a small contact region of length 
$2L_x+1$ and maximum width $2L_y+1$. $c_{{\bf i}\sigma}$ ($c^{\dagger}_{{\bf i}\sigma}$) being the destruction (creation) operator of an 
electron of spin $\sigma$ at site ${\bf i}$ of coordinates $(i_x,i_y)$, and $n_{{\bf i}\sigma}=c^{\dagger}_{{\bf i}\sigma}c_{{\bf i}\sigma}$, 
the Hamiltonians of the left ($i_x\leq -L_x$) and right ($i_x\geq L_x$) leads read
\begin{equation}
H_{leads}= \sum_{{\bf i},\sigma} \left( -4t n_{{\bf i}\sigma} + t \sum_{\bf j} c^{\dagger}_{{\bf i}\sigma} c_{{\bf j}\sigma} \right) + H.C. 
\label{lead-Hamiltonian}
\end{equation}
The hopping terms are non-zero between nearest neighbors sites ${\bf i,j}$ only. The energy scale is 
defined by taking $t=-1$ (the conduction bands of the leads are in the energy interval $[0,8]$ when the site 
potentials are equal to $-4t$). Hereafter, we study the continuum limit (energy $E \ll 1$). The contact Hamiltonian reads
\begin{equation}
H_{contact}= \sum_{{\bf i},\sigma} \left( (V_{\bf i}-4t) n_{{\bf i}\sigma} + t \sum_{\bf j} c^{\dagger}_{{\bf i}\sigma} 
c_{{\bf j}\sigma} \right) + H.C. 
\label{contact-Hamiltonian}
\end{equation}
The summations are restricted to $-L_x\leq i_x \leq L_x$ and to $-L_y\leq i_y \leq L_y$. Moreover, the site potentials $V_{\bf i}$ are 
taken infinite inside the contact region if $|i_y| \geq (L_y-k)+k \left(i_x/L_x \right)^2$, where $k$ is a parameter. This restricts 
the electron motion inside a smoothly opening region known to favor a sharp opening of the conductance channels as one increases the 
energy. A smooth opening also reduces the interference effects induced by the back-scattering of electron waves leaving the contact region, 
effects which induce oscillations in the transmission function $T(E)$. The Hamiltonians describing the coupling between the contact and the 
two leads read
\begin{eqnarray}
H_c^l &=& t_c \sum_{i_y=-L_y,\sigma}^{L_y}  \left(c^{\dagger}_{{(-L_x,i_y)}\sigma} c_{{(-L_x-1,i_y)}\sigma}+H.C.\right), \\
H_c^r &=& t_c \sum_{i_y=-L_y,\sigma}^{L_y}  \left(c^{\dagger}_{{(L_x,i_y)}\sigma} c_{{(L_x+1,i_y)}\sigma}+H.C.\right).
 \label{coupling-Hamiltonian}
\end{eqnarray}
The QPC Hamiltonian reads $H_{0}=H_{contact}+\sum_{\alpha=l,r} (H_c^{\alpha}+H_{leads}^{\alpha})$. The QPC transmission $T_0(E)$ is 
a staircase function, each stair taking an integer value which counts the number of open channels. To describe the depletion region 
induced by the charged tip, a term $H_{tip}(x,y)=\sum_{\sigma} V n_{{\bf T}\sigma}$ is added to $H_{0}$, which modifies by an amount $V$ 
the potential $-4t$ of a single site ${\bf T}$ of coordinates $(x,y)$ located at a distance $r=\sqrt{x^2+y^2}$ from the contact. 
The interferometer Hamiltonian reads $H=H_{0}+H_{tip}(x,y)$. Fig.~\ref{fig2QPC} shows such an interferometer when $L_x=L_y=3$ and $k=2$.

\begin{figure}
\includegraphics[keepaspectratio,width=\columnwidth]{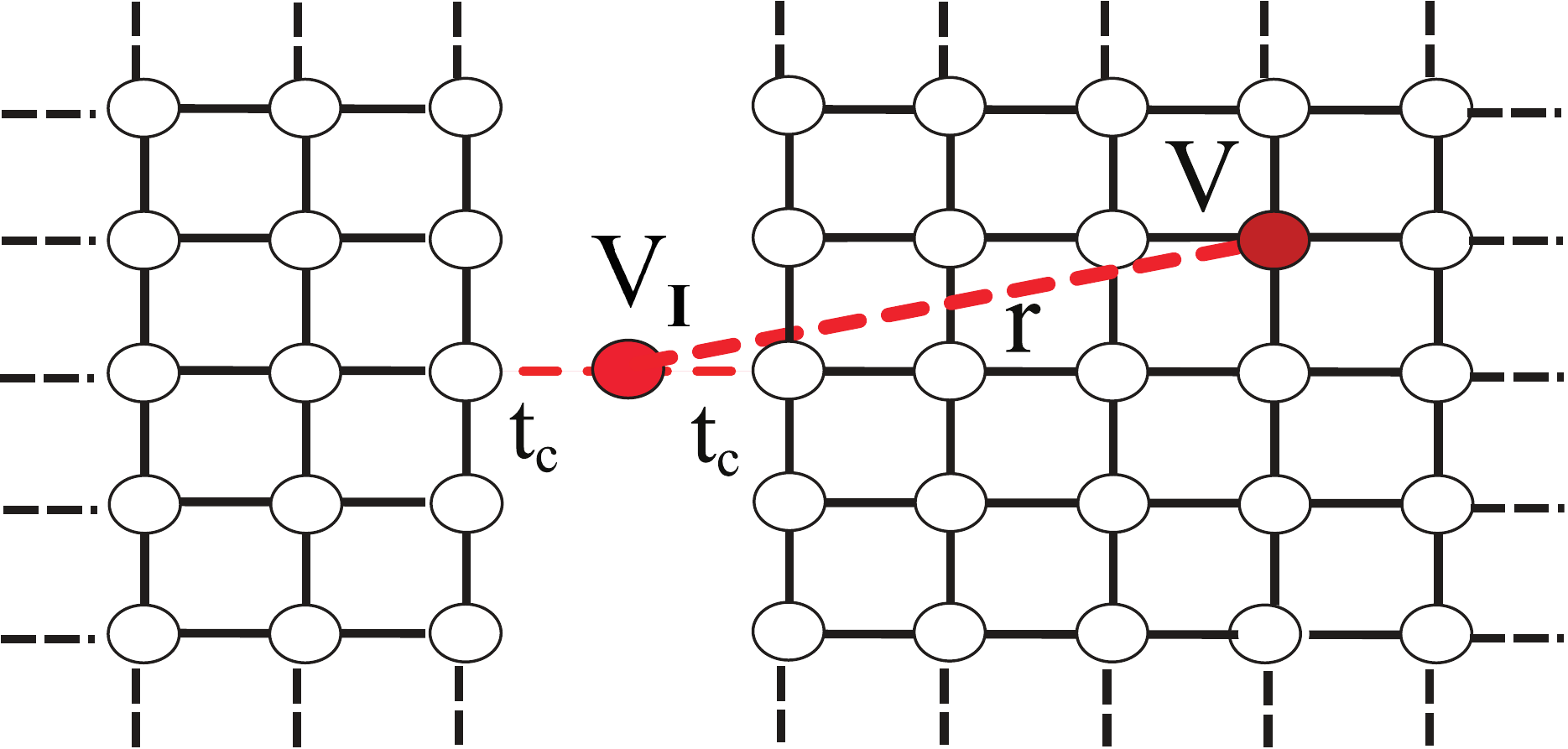}
\caption{(Color online) RCM model for the SGM of a quantum contact: Two semi-infinite square lattices are contacted via a single 
site ${\bf I}$ of coordinates $(0,0)$, potential $-4t+V_{\bf I}$ and coupling term $t_c$. Taking a potential $-4t+V$ at another 
single site ${\bf T}$ of coordinates $(x,y)$ gives rise to an electron interferometer of size $r=\sqrt{x^2+y^2}$.} 
\label{fig2RCM}
\end{figure}

Before studying a QPC when $L_x$ and $L_y$ are large in section~\ref{QPC}, it is very instructive to study the limit shown 
in Fig.~\ref{fig2RCM} where the contact is reduced to a single site ${\bf I}$ of coordinates $(0,0)$ and potential $-4t+V_{\bf I}$. 
This defines the resonant contact model (RCM) which can be solved analytically~\cite{alp} when the width of the leads becomes infinite. 
As one varies the energy $E$ for $t_c \ll 1$ and $V=0$, the RCM transmission $T_0(E)$ exhibits a single spin-degenerate Breit-Wigner 
resonance, and not the usual staircase function which characterizes the QPC conductance quantization.

\section{SGM of the Resonant Contact Model}
 
\subsection{RCM model: Spin-degenerate case at ${\cal T}=0$}

For the RCM contact without tip, the transmission of an electron of spin $\sigma$ and energy $E$ is given by 
the Fisher-Lee formula \cite{Datta:book97}:
\begin{equation}\label{eq:FL}
T_0^{\sigma}(E) = \text{Tr} \left [\Gamma_l(E)  G_{0}^R(E)  \Gamma_r(E) G_{0}^A(E)  \right ] \; , 
\end{equation}
where $G_{0}^R$ is the retarded Green's function of the contact dressed by the right (r) and left (l) leads:
\begin{equation}\label{eq:GF0}
 G_{0}^R (E) = \underset{\eta \rightarrow 0^+}{\text{lim}} (E+i \eta - 4 - V_{\bf I} - \Sigma_l^R - \Sigma_r^R)^{-1} \; . 
\end{equation}
The contact being reduced to a single site ${\bf I}$ coupled to another single site per lead, the lead self-energies 
$\Sigma_{l,r}(E)$ are only two complex numbers $\Sigma_{l,r}(E)=R_{l,r}(E)+iI_{l,r}(E)=t_c^2<\pm 1,0|G_{l,r}^R(E)|\pm 1,0>$ 
when $t=-1$ and where $G^R_{l,r}(E)$ are the retarded Green's function of the left and right leads evaluated at the sites 
directly coupled to ${\bf I}$. The coupling rates to the right and left leads verify: $\Gamma_{r,l} =  i (\Sigma_{r,l}^R - \Sigma_{r,l}^A )$. 
Using the method of mirror images~\cite{Molina:PRB06}, $G^R_{l,r}(E)$ can be expressed in terms of the Green's function 
$G^R_{2D}(E)$ of the infinite 2D square lattice~\cite{Economou:book06}. One gets: 
\begin{equation}
T_0^{\sigma}(E)= \frac{4 I_rI_l}{(E-4-V_{\bf I}-R_r-R_l)^2+(I_r+I_l)^2}.     
\label{transmission_without_tip}
\end{equation}
If the variation of $\Sigma_{l,r}(E)$ can be neglected when $E$ varies inside the resonance (typically $t_c<0.5$ in the continuum limit 
where the Fermi momentum $k_F \ll 1$), one gets a Lorentzian of width $\Gamma = -2I$ and center $4+V_{\bf I}+2R$ since $R_l=R_r \equiv R$ 
and $I_r=I_l\equiv I$.\\ 
\indent If one adds a tip potential $V \neq 0$ in the right lead, the effect of the tip can be included by adding an amount 
$\Delta \Sigma_r(E)=\Delta R_{r}(E)+i\Delta I_{r}(E)$ to $\Sigma_r(E)$. The interferometer transmission $T^{\sigma}(E)$ is still given 
by Eq.~\eqref{transmission_without_tip}, once $R_r+\Delta R_r$ and $I_r+\Delta I_r$ have been substituted for $R_r$ and  $I_r$ 
(see Refs.~\cite{alp,Darancet:PRB10}). When the effect of the charged tip is restricted to a single site ${\bf T}$, 
$\Delta \Sigma_r$ can be obtained from Dyson's equation for $G_{r+V}(E)$ the Green's function of the right lead with the tip potential:
\begin{eqnarray}
<1,0| G_{r+V}^R(E)|1,0>= <1,0| G_{r}^R(E)|1,0> \nonumber \\
+ \frac{<1,0| G_{r}^R(E)|{\bf T}> V <{\bf T}| G_{r}^R(E)|1,0>}{1-V <{\bf T}| G_{r}^R(E)|{\bf T}>}
\end{eqnarray}
In the continuum limit and for distances $r \gg k_F^{-1}$, one finds:
\begin{equation}   
\frac{\Delta \Sigma_r}{t_c^2 \rho} \approx -\frac{k_F x^2}{2 \pi r^3} \exp[i(2k_Fr+\pi/2+\phi)] + O(\frac{x^{3/2}}{r^3}),   
\label{self-energy}
\end{equation} 
where $\rho$ and $\phi$ are the modulus and the phase of the amplitude of $V/(1-V \langle 0,0|G^R_{2D}(E)|0,0\rangle )$. 
In the continuum limit, $I  \approx -t_c^2 k_F^2/4$. Therefore, at sufficiently large distances 
$r \gg \lambda_F/2$, $\Delta \Sigma_r^R(E) \ll I$ and one can expand $T^{\sigma}(E)$ to the leading order $\propto x^2/r^3$ in $\Delta \Sigma_r$:
\begin{eqnarray}
 \dfrac{T^{\sigma}(E)-T^{\sigma}_{0}(E)}{T^{\sigma}_{0}(E)} \approx &  -s \sqrt{T^{\sigma}_{0}(E) (1-T^{\sigma}_{0}(E))} \; \frac{\Delta R_r(E)}{I} \nonumber \\
 & + (1- T^{\sigma}_{0}(E)) \; \frac{\Delta I_r(E)}{I} \; ,
\end{eqnarray}
where $s= \text{sign}[V_{\bf I}^{res}-V_{\bf I}]$ and $V^{res}_{\bf I}\equiv E-4-2R$ is the value of $V_{\bf I}$ where 
$T^{\uparrow}_{0}=T^{\downarrow}_{0}=1$. This leads to the simple prediction:
\begin{eqnarray}
 \dfrac{T^{\sigma}(E)-T^{\sigma}_{0}(E)}{T^{\sigma}_{0}(E)} \approx A_0  \cos (2 k_F r + \Phi_0)  + O\left( \dfrac{x^{3/2}}{{r}^{3}}\right)\; 
\label{SGM-T=0}
\end{eqnarray}
where the amplitude $A_0=\frac{2 {\rho}}{\pi k_F} \frac{x^2}{r^3}  \sin \zeta_0 $ decreases as $\frac{x^2}{r^3}$, the 
phase $\Phi_0= \pi/2 + {\phi} - \zeta_0$ and $\sin \zeta_0 = - s \sqrt{1- T^{\sigma}_{0}(E)}$. Eq.~\eqref{SGM-T=0} describes  
Fabry-P\'erot fringes spaced by $\lambda_F/2$ and their decay with $r$, assuming $T^{\sigma}_{0}(E)<1$. One needs to take into account 
corrections of higher order~\cite{alp} when $T^{\sigma}_{0}(E) \to 1$, a limit which we will not consider in this work.

\subsection{RCM model: Spin-degenerate case at ${\cal T} \neq 0$}
Let us now study how the effect of the tip upon the conductance $g$ in units of $e^2/h$ depends on the temperature ${\cal T}$:
\begin{equation}
\Delta g=g-g_0= \sum_{\sigma}\int dE  (T^{\sigma}(E)-T^{\sigma}_{0}(E)) (-\frac{\partial f}{\partial E}),
\label{conductance-change}
\end{equation} 
where $f$ is the Fermi-Dirac distribution. Let us consider the case of a sharp Lorentzian resonance ($t_c <0.5$) of $T^{\sigma}_{0}(E)$. 
Then, $R$ and $I$ do not vary rapidly inside the resonance. In the same way, $\rho$ and $\phi$ vary slowly inside the resonance and thus 
can be considered as constants. In order to calculate the integral \eqref{conductance-change} analytically, we make the approximation~\cite{Heller:NL05} 
$-\partial f/ \partial E \approx (1/4k_B{\cal T}) \exp-[\sqrt{\pi}(E-E_F)/(4k_B{\cal T})]^2$, where $E_F$ and $k_B$ are 
the Fermi energy and the Boltzmann constant. One gets $\Delta g = D_1+ D_2$ where: 
\begin{widetext}
\begin{eqnarray}\label{eq:Temp-D1-D2-hors-res}
 D_1 &\approx& \frac{2 \rho}{\pi^{3/2} k_F } \dfrac{x^2}{r^3}\frac{l_{\cal T}}{l_\Gamma} \;\; \Re \left[  e^{i (2 k_F r+ \phi+\pi/2)} 
\int_{-\infty}^{\infty} \dfrac{q+v}{\left[1+(q+v)^2\right]^2} e^{-\left(\frac{l_{\cal T}}{l_\Gamma} q\right)^2} e^{i \frac{r}{l_\Gamma} q} \; \ud q \right] \; , \\
 D_2 &\approx & \frac{2 \rho_T}{\pi^{3/2} k_F } \dfrac{x^2}{r^3}\frac{l_{\cal T}}{l_\Gamma} \;\; \Im \left[  e^{i (2 k_F r+ \phi_T+\pi/2)} 
\int_{-\infty}^{\infty} \dfrac{(q+v)^2}{\left[1+(q+v)^2\right]^2} e^{-\left(\frac{l_{\cal T}}{l_\Gamma} q\right)^2} e^{i \frac{r}{l_\Gamma} q} \; \ud q \right] \; .
\end{eqnarray}
\end{widetext}
$v \equiv (V^{res}_{\bf I}-V_{\bf I})/\Gamma$ gives the energy shift of $V_{\bf I}$ from the resonance $V^{res}_{\bf I}$ in units of 
$\Gamma$, and $q=(E-E_F)/\Gamma$. $l_{\cal T}$ and $l_{\Gamma}$ are the two length scales respectively associated to ${\cal T}$ (Fermi-Dirac 
statistics) and to $\Gamma$ (resonant transmission):
\begin{eqnarray}
\label{scales}
l_{\cal T}&=&\frac{\sqrt{\pi}k_F}{4k_B{\cal T}} \\
l_{\Gamma}&=& \frac{k_F}{\Gamma}.
\end{eqnarray}
Calculating the Fourier transforms $D_1$ and $D_2$, we eventually obtain:
\begin{equation}\label{eq:deltaghorsres}
\Delta g ({\cal T})\approx 2 A({\cal T}) \cos (2k_Fr+\Phi({\cal T})) \;, 
\end{equation}
where at large distance $r > r^*\equiv 2 l_{\cal T}[1+l_{\cal T}(1+|v|)/l_{\Gamma}]$:
\begin{equation}
A({\cal T})=\frac{\rho x^2 l_{{\cal T}}}{\sqrt{\pi}k_F r^3} \exp -[(1+v^2)(\frac{l_{\cal T}}{l_{\Gamma}})^2 +\frac{r}{l_{\Gamma}}] 
\label{conductance1}
\end{equation}
\begin{equation}
\Phi({\cal T})=\phi+v\frac{r}{l_{\Gamma}}-2v(\frac{l_{\cal T}}{l_{\Gamma}})^2. 
\label{conductance2}
\end{equation}
The factor $2$ in $\Delta g ({\cal T})$ comes from the spin degeneracy. 

\subsection{RCM model: Effect of a parallel magnetic field applied everywhere}
\begin{figure}
\includegraphics[keepaspectratio,width=\columnwidth]{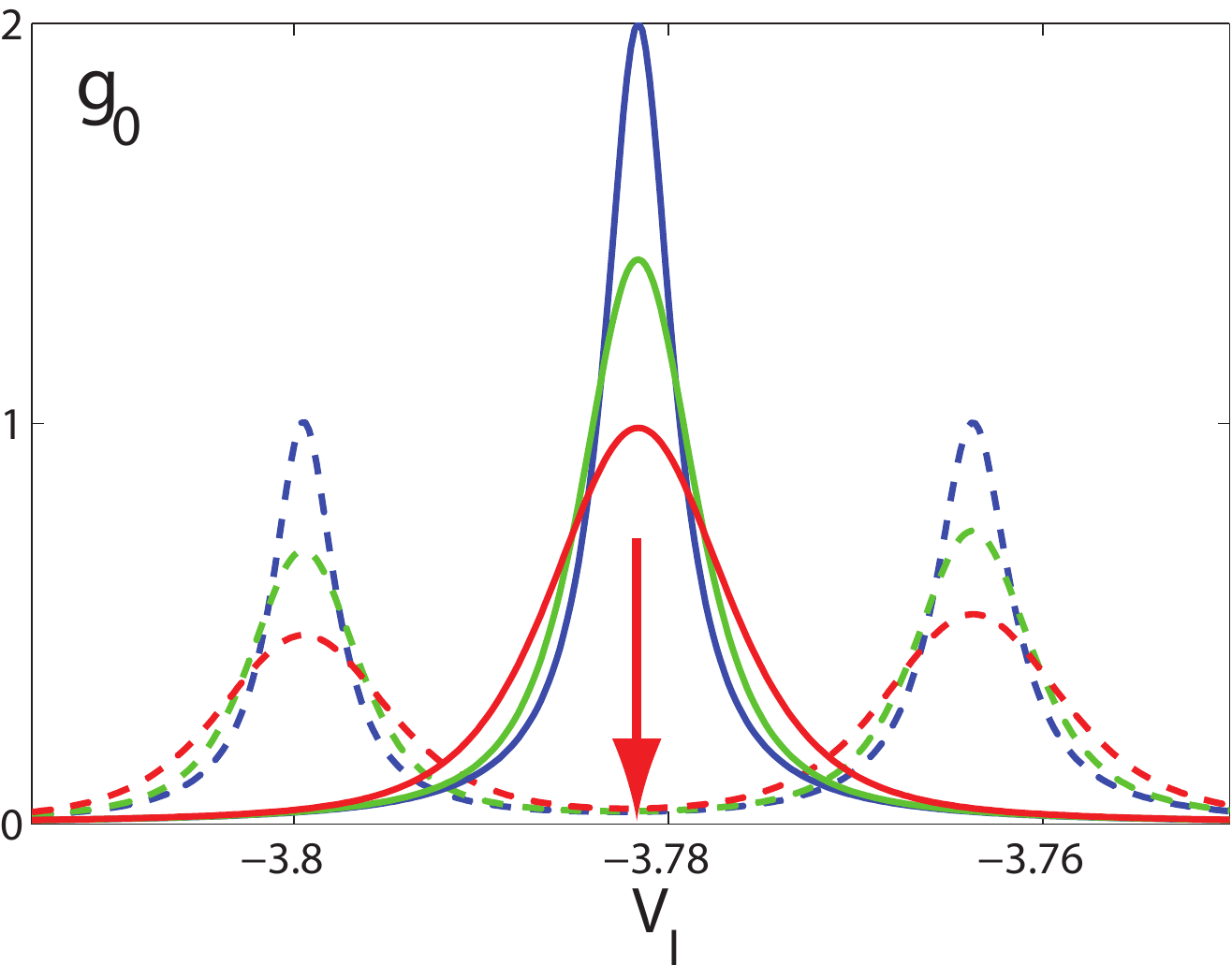}
\caption{(Color online) Conductance $g_0$ (in units of $e^2/h$) of the RCM model without field ($h=0$, solid line) and 
with a uniform parallel magnetic field ($h=8\Gamma$, dashed line) as a function of $V_{\bf I}$ for ${\cal T}=0$ (blue) 
${\cal T}=\Gamma/2$ (green) and ${\cal T}=\Gamma$ (red). There is no tip, $\Gamma=0.003$ ($t_c=0.2$ and $E_F=0.15$) 
and the field is applied everywhere. The arrow gives the symmetric point where the SGM images are studied.}
\label{fig4}
\end{figure}

 Let us consider first the case where a uniform parallel magnetic field is applied everywhere. The spin degeneracy is broken 
and the electrons of opposite spin have energies which are shifted ($E^{\sigma} \to E(h=0)\pm h$). They contribute to transport 
with different wave-vectors in the 2DEG and a somewhat trivial beating effect is induced in the SGM images, the Fabry-P\'erot fringes 
having different wavelengths ($\lambda^{\uparrow} \neq \lambda^{\downarrow}$). When $E_F$ is small enough, we can use the continuum dispersion 
relation $k^{\sigma}=\sqrt{E^{\sigma}}$. If the Zeeman energy $h$ remains small compared to $E_F$, the 2DEG is not fully polarized. In 
Fig.~\ref{fig4}, the conductance $g_0$ of the RCM contact is given as a function of the contact potential $V_I$. One can see how the spin 
degenerate peak of conductance ($h=0$) of width $\Gamma$ is split by the field ($h=8 \Gamma$) for increasing values of ${\cal T}$. 
Hereafter, we study the SGM image when the contact is open between the peaks  (symmetric point indicated by an arrow in Fig.~\ref{fig4} 
where $V_{\bf I}=V_{\bf I}^{res}(h=0)$). 

\begin{figure}
\includegraphics[keepaspectratio,width=\columnwidth]{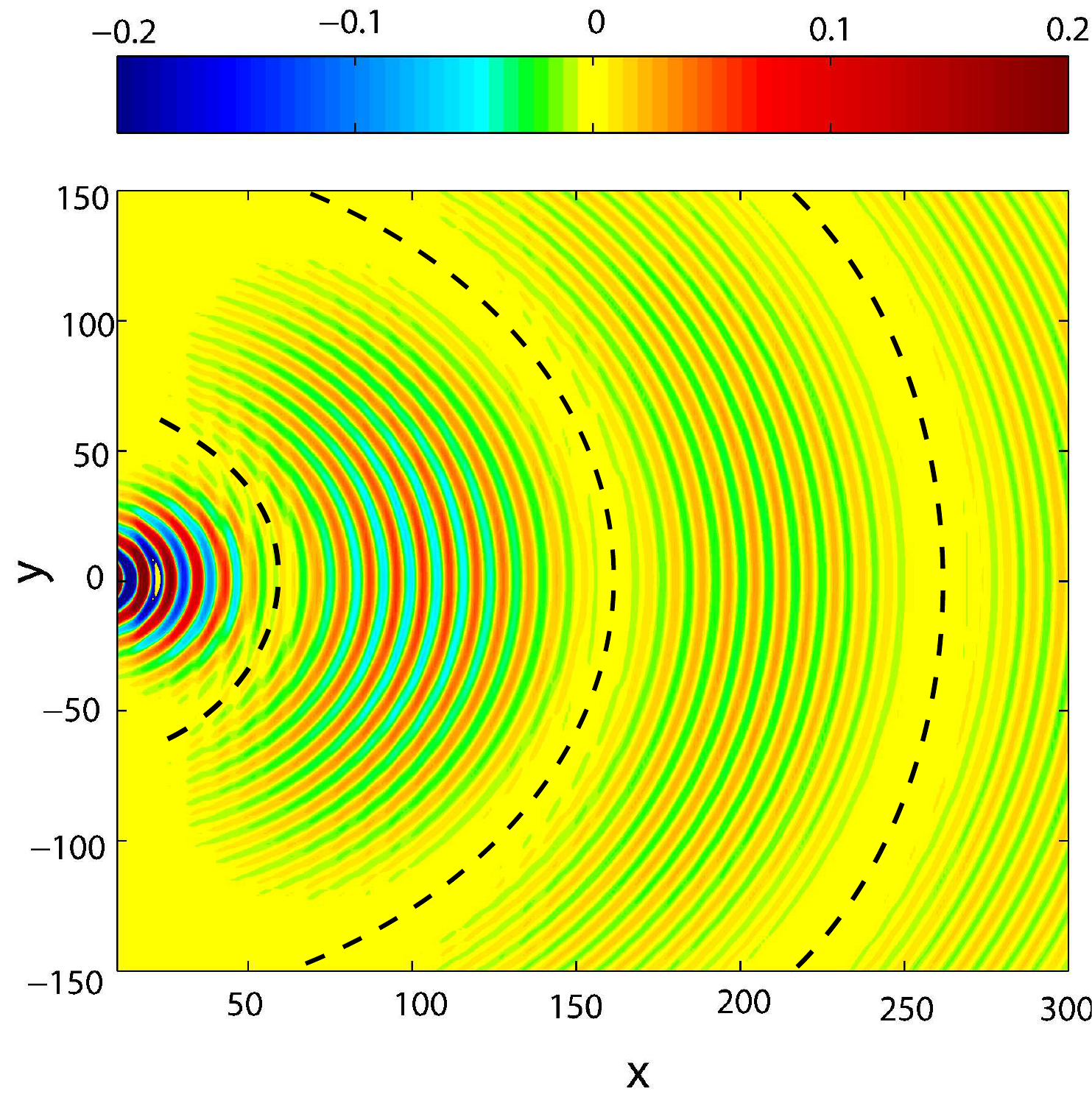}
\caption{(Color online) (Color online) RCM model with a parallel magnetic field applied everywhere ($h=4\Gamma\approx 0.0119$) at a temperature 
${\cal T}=0$. The relative effect $\Delta T/T_0$ (upper color scale) of the tip ($V=-2$) upon the RCM transmission $T_0$ has been 
numerically calculated and is given as a function of the tip coordinates $(x,y)$ when $t_c=0.2$ and $E_F=0.15$. The contact potential 
has the value $V_I$ indicated by the arrow in Fig.~\ref{fig4} ($T_0=0.12$ and $\Gamma\approx 0.003$). The dashed lines give the rings 
of radii $r^{\cal D}(n)$ predicted by Eq.~\eqref{radii}.}
\label{fig5}
\end{figure}

The relative effect of the tip upon the transmission of an electron of spin $\sigma$ at an energy $E$ becomes, 
\begin{eqnarray}
\dfrac{T^{\sigma}(E)-T^{\sigma}_{0}(E)}{T^{\sigma}_{0}(E)} \approx A_0^{\sigma}  \cos (2 k_F^{\sigma} r + \Phi_0^{\sigma})  
+ O\left( \dfrac{x^{3/2}}{{r}^{3}}\right)\; 
\label{deltaT}
\end{eqnarray}
where $A_0^{\sigma}= (2 {\rho}^{\sigma})/(\pi k_F^{\sigma}) (x^2/r^3)  \sin \zeta_0^{\sigma}$, 
$\Phi_0^{\sigma}= \pi/2 + {\phi}^{\sigma} - \zeta_0^{\sigma}$, 
$\sin \zeta_0^{\sigma} = - s^{\sigma} \sqrt{1- T^{\sigma}_{0}(E)}$ and $s^{\sigma}=\text{sign}(V_I^{res}-V_I)$.
If the contact is opened in the middle between the two transmission peaks, $s^{\uparrow}=-s^{\downarrow}$, 
$\sin \zeta_0^{\uparrow}=-\sin \zeta_0^{\downarrow}$. The contribution of electrons of opposite spins to 
$\sum_{\sigma} (T^{\sigma}(E)-T^{\sigma}_{0}(E))/T^{\sigma}_{0}(E)$ have opposite signs when the tip is put at a distance
\begin{eqnarray}
r^{\cal D}(n)=\frac{2\pi n +({\phi}^{\downarrow}-{\phi}^{\uparrow})+({\rho}^{\uparrow}-{\rho}^{\downarrow})}{2(k_F^{\uparrow}-k_F^{\downarrow})}\;,   
\end{eqnarray}
where $n$ is integer ($0,1,2,\ldots$).
At a temperature ${\cal T}=0$, this means that the SGM image of a contact opened between its two transmission resonances exhibits a pattern 
of rings of radii $r^{\cal D}$, where the beating between the contribution of opposite spin is destructive. In contrast, the beating becomes 
constructive on rings of radii $r^{\cal C}$. Neglecting the small spin dependence of $T^{\sigma}_{0}$ at the symmetric point, these radii 
become independent of the temperature ${\cal T}$ and reads
\begin{equation}
r^{\cal D}(n) \approx \frac{\pi n + \arcsin (\sqrt{1-T_0})}{\sqrt{E_F+h}-\sqrt{E_F-h}} \\
\label{radii}
\end{equation}
\begin{equation}
r^{\cal C}(n) \approx \frac{(n+1/2)\pi+ \arcsin (\sqrt{1-T_0})}{\sqrt{E_F+h}-\sqrt{E_F-h}}\;.
\label{radii-c}
\end{equation}
 In Fig.~\ref{fig5}, one can see a SGM image taken with a parallel magnetic field at ${\cal T}=0$: We can see the three first rings 
at the expected radii $r^{\cal D}(n)$ with $n=0,1,2$ (dashed lines), where the effect of the tip upon $g$ is suppressed, separated 
by regions centered around rings of radii $r^{\cal C}(n)$ where this effect is enhanced by the applied magnetic field.    

\begin{figure}
\includegraphics[keepaspectratio,width=\columnwidth]{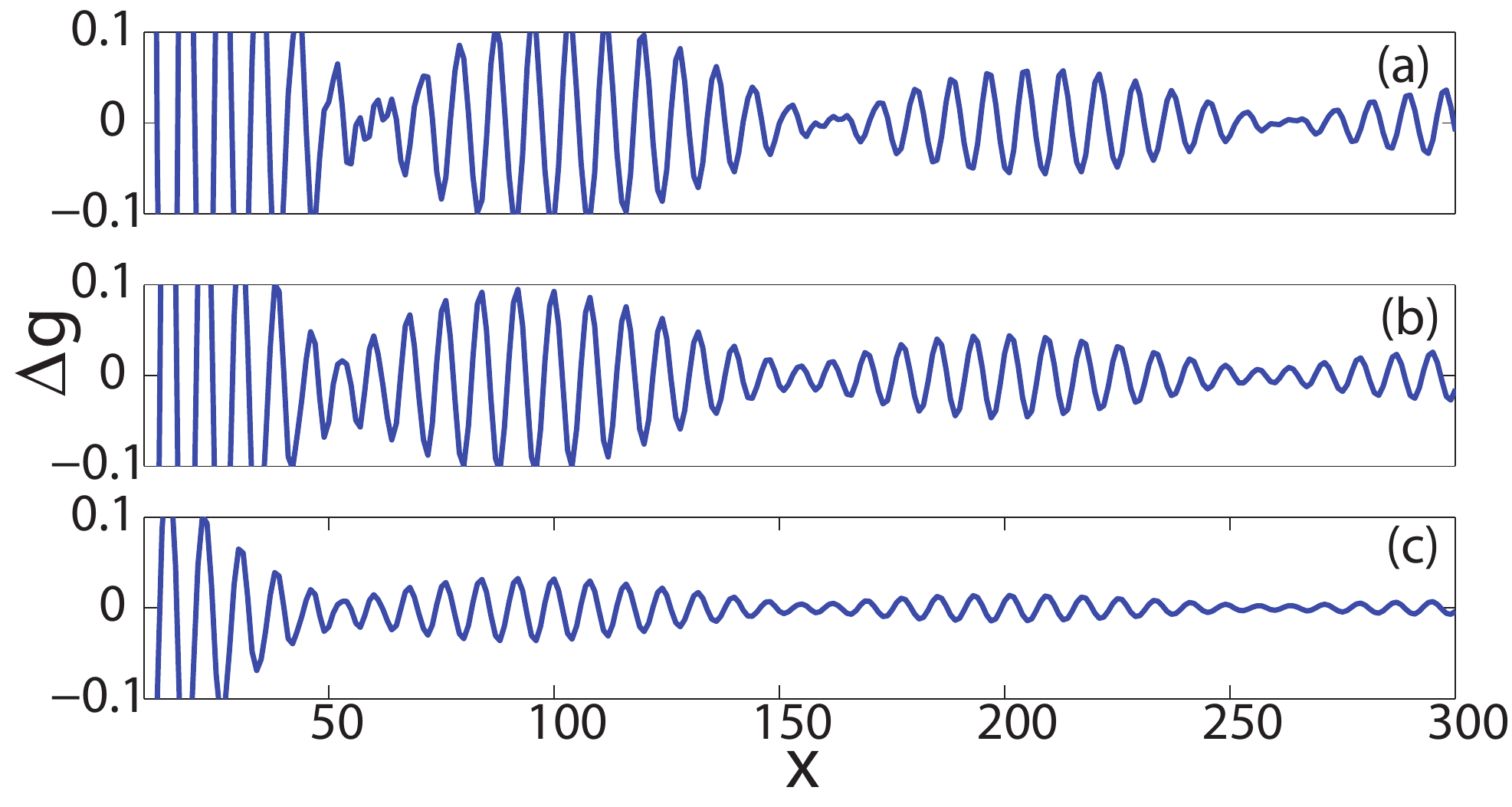}
\caption{(Color online) 
RCM contact with a parallel magnetic field applied everywhere ($h=4\Gamma\approx 0.0119$): The numerical 
values of $\Delta g$ are given as a function of the tip coordinates $(x,y=0)$ for $E_F=0.15$, $V=-2$, $t_c=0.2$ and 
$\Gamma\approx 0.003$. The figures (a) ${\cal T}=0$, (b) $k_B{\cal T}=10^{-4}$, (c) $k_B{\cal T}=8\times 10^{-4}$ are in the 
regime $k_B{\cal T}/{\Gamma} \ll v$. The location $r^{\cal D}(n)$ of the destructive interferences is independent of the temperature 
${\cal T}$, in agreement with Eq.~\eqref{deltagwithT}.}
\label{fig6}
\end{figure}

When the temperature ${\cal T} \neq 0$ but satisfies the condition $k_B{\cal T}/\Gamma \ll v$, the expressions can be simplified 
if $r \ll 2 l_{\cal T} \left[1+(l_{\cal T}/l_{\Gamma})(1+|v|)\right]$. One finds that the SGM images are roughly identical to those 
described by Eq.~\eqref{deltaT} for ${\cal T}=0$ within a circle of radius $l_{\cal T}$, and are suppressed outside (see Fig.~\ref{fig6})
\begin{equation}
\label{deltagwithT}
\Delta g ({\cal T},h) \approx  \Delta g ({\cal T}=0,h) \exp-(\frac{r}{2l_{\cal T}})^2 \;. 
\end{equation}
\subsection{RCM model: Effect of a parallel magnetic field applied upon the contact}
\begin{figure}
\includegraphics[keepaspectratio,width=\columnwidth]{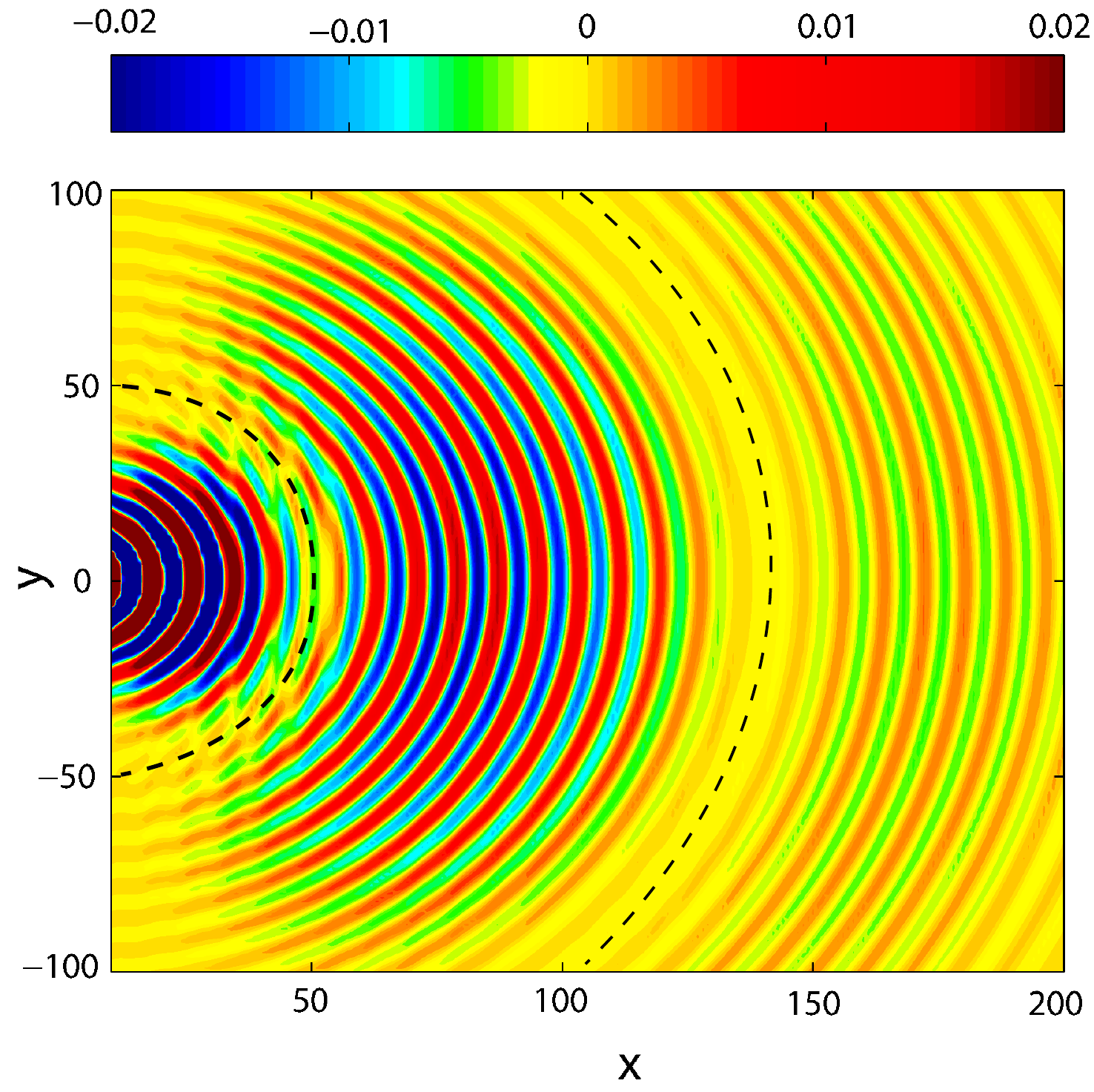}
\caption{(Color online)  
RCM model with a Zeeman term $h= \pm 0.0136$ in the contact only at a temperature ${\cal T}^*=0.0099/k_B$ ($l_{\cal T}=17.34$).
The numerically calculated values of  $\Delta g/g_0 ({\cal T}^*,\Gamma^*)$ are plotted as a function of the coordinates $(x,y)$ 
of the tip (potential $V=-2$). $\Gamma^*=0.0035$. The parameters have been chosen such that the radius (Eq.~\eqref{radius-ring}) of 
the first ring $r^D(n=0)=50$. The dashed lines give the circles of radii $r^{\cal D}(n)$ predicted by the theory (Eq.~\ref{radius-ring}).}
\label{fig8}
\end{figure}

Let us consider now the case where a parallel magnetic field is applied upon the contact only. 
This removes the spin degeneracy in the contact by a local Zeeman term $\pm h$. 
In contrast to the previous case, $k_F$ and hence $l_{\cal T}$ and $l_{\Gamma}$ remain independent of $\sigma$, while $\Delta \phi=0$. 
There is nevertheless a beating effect between the interference fringes of opposite spins, which exhibits a more unusual 
temperature dependence than before: It can be observed only when the temperature becomes of the order of the Zeeman splitting, 
but vanishes as ${\cal T}\to 0$.\\
\indent Let us consider the value of $V_{\bf I}$ (symmetric point) where there is the resonance without field 
(total transmission $T_0(h=0)=2$), one has $v_{\uparrow}=-v_{\downarrow}$ and Eq.~\eqref{eq:deltaghorsres} gives 
\begin{equation} 
r^{\cal D}(n)=\frac{2k_F}{\Gamma}(\frac{l_{\cal T}}{l_{\Gamma}})^2+(n+\frac{1}{2}) \frac{\pi k_F}{h} 
\label{radius-ring}
\end{equation}   
($n=0,1,\ldots$) for the radii of the rings where the effect of the tip is suppressed by the field. Conversely, the oscillations of  
$\Delta g^{\uparrow}({\cal T})$ and $\Delta g^{\downarrow}({\cal T})$ add if the distance $r$ is given by $r^{\cal C}(n)=r^{\cal D}(n)+\pi k_F/(2h)$. 
The SGM image is characterized by a first ring at a distance $r^{\cal D}(n=0)$ followed by other rings spaced by $\pi k_F/h$ where 
$\Delta g ({\cal T})=0$. To optimize the contrast in the images, we calculate for a given value of $h$ the temperature ${\cal T}^*$ and the width 
$\Gamma^*$ for which $\Delta g({\cal T},r=r^{\cal C}(n=0))$ is maximum. The extrema are given by the conditions $\partial A / \partial l_{\Gamma}=0$ 
and $\partial A / \partial l_{\cal T} =0$. This gives two coupled non-linear algebraic equations which can be solved numerically, yielding 
\begin{equation}
k_B{\cal T}^* \approx 0.73 h \ \ \ \Gamma^*\approx 0.25 h.  
\end{equation} 
To observe the rings, their spacing must exceed $\lambda_F$ ($h<(\pi/\lambda_F)^2$). In Fig.~\ref{fig8}, a numerical calculation 
of an SGM image is shown when ${\cal T}={\cal T}^*$ and $\Gamma=\Gamma^*$. In the presence of a Zeeman term $h$ in the contact, 
the image exhibits the ring pattern predicted by the theory.
\begin{figure}
\includegraphics[keepaspectratio,width=\columnwidth]{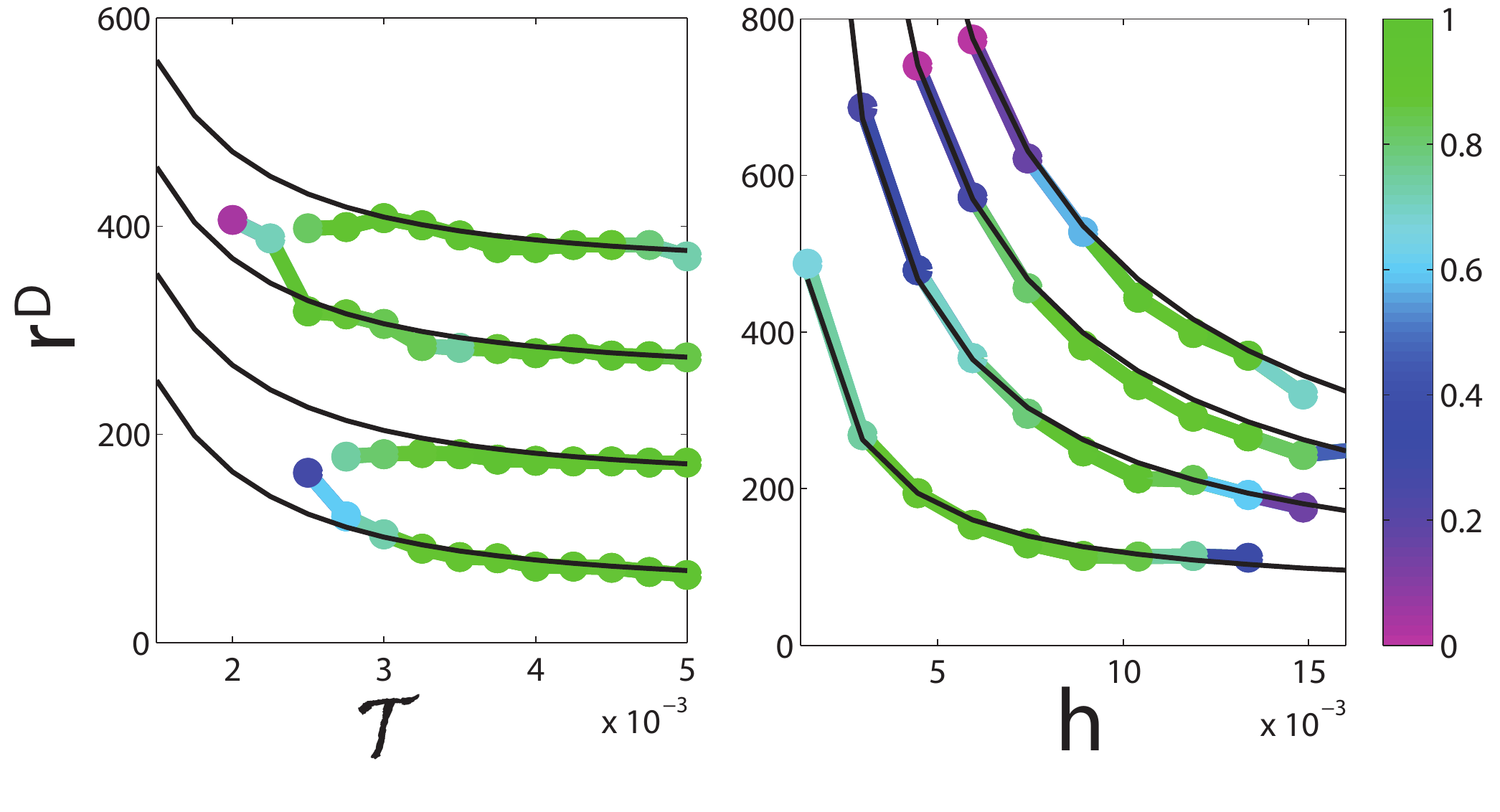}
\caption{(Color online) 
Radii $r^{\cal D}(n)$ of the successive rings ($n=0, 1, 2$ and $3$) when a local field $h$ is applied in the contact and 
$\Gamma=0.003$: Left figure: Radii $r^{\cal D}(n)$ as a function of $k_B{\cal T}$ when $h=0.01$. Right figure: $r^{\cal D}(n)$ as a function of 
$h$ when $k_B{\cal T}=0.0028$. The dots give the successive radii where the numerically calculated values of $\Delta g/g_0 \approx 0$, their colors 
corresponding to a visibility scale indicated at the right ($0$ without contrast, $1$ for the best contrast). The solid lines give the analytical 
expression~\eqref{radius-ring} of the radii $r^{\cal D}(n)$ which was derived assuming $r^{\cal D}(n)>r^*$.} 
\label{fig11}
\end{figure}
Fig.~\ref{fig11} shows us how the radii $r^{\cal D}(n)$ of the successive rings where the tip does not change the conductance depend on the 
temperature ${\cal T}$ for a given local field $h$, or on the field $h$ for a given temperature ${\cal T}$. For a resonance 
width $\Gamma=0.003$, one can see how the $r^{\cal D}(n)$ increase when ${\cal T} \to 0$ or $h \to 0$, making impossible the observation 
of a beating effect in those limits. In Fig.~\ref{fig9}, the relative change $\Delta g(x,y=0)/g_0$ of the conductance is shown in the presence 
of a Zeeman term in the contact as one varies the tip coordinate $x$ (keeping $y=0$). This change is given for three different values of $h$, 
when the temperature ${\cal T}$ and the resonance width $\Gamma$ take their optimal values ${\cal T}^*$ and $\Gamma^*$. 
\begin{figure}
\includegraphics[keepaspectratio,width=\columnwidth]{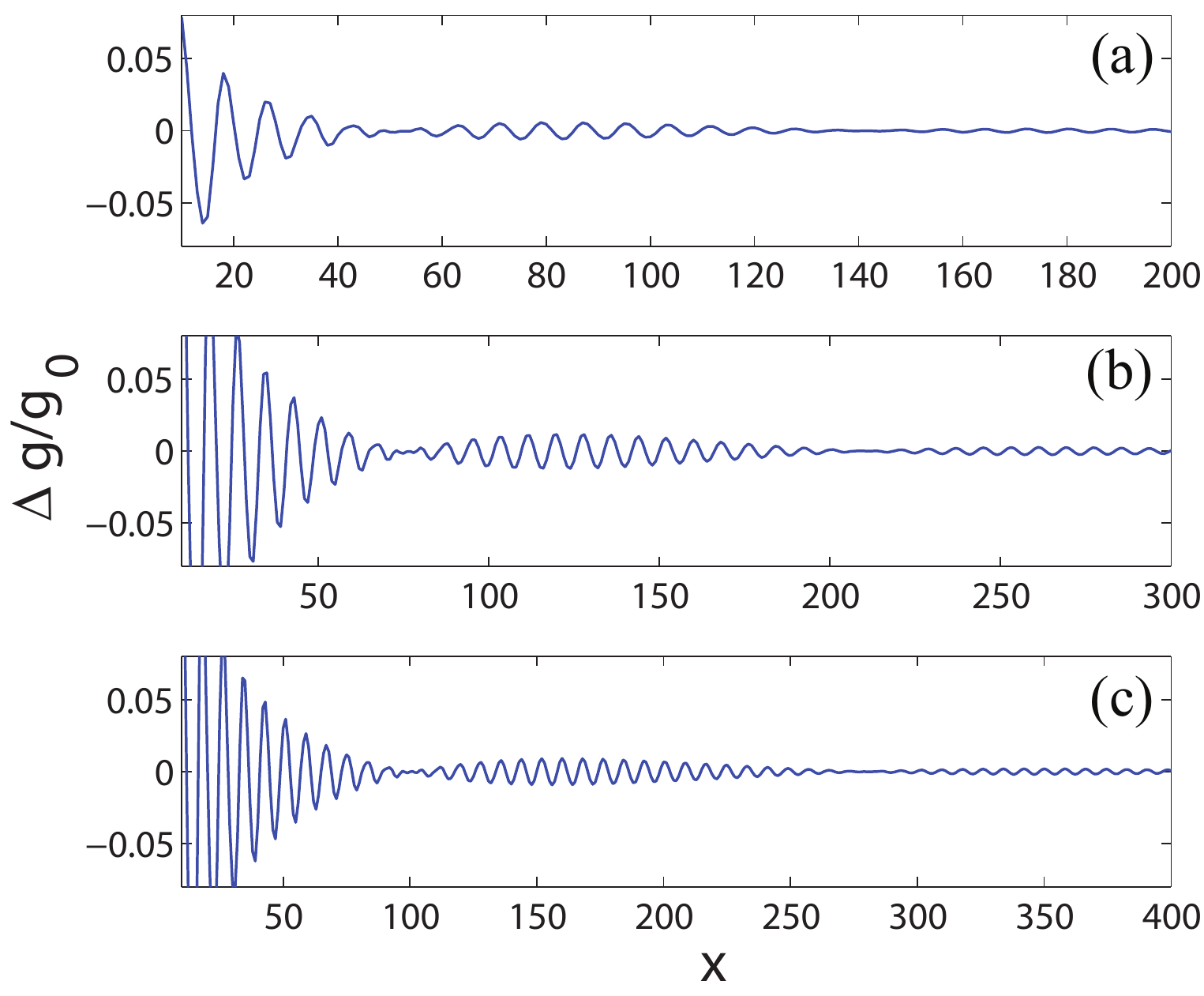}
\caption{(Color online)  $\Delta g/g_0 ({\cal T}^*,\Gamma^*,y=0)$ as a function of tip coordinate $x$ (keeping $y=0$) for three values $h$ 
of a Zeeman term added in the RCM contact only: $V=-2$, $E_F=0.1542$ and $\lambda_F/2=8$. (a): $h=0.0136$, $k_B{\cal T}^*=0.0099$, $\Gamma^*=0.0035$ 
(b): $h=0.0091$, $k_B{\cal T}^*=0.0066$, $\Gamma^*=0.0023$; (c): $h=0.0068$, $k_B{\cal T}^*=0.005$, $\Gamma^*=0.0017$.} 
\label{fig9}
\end{figure}
\begin{figure}
\includegraphics[keepaspectratio,width=\columnwidth]{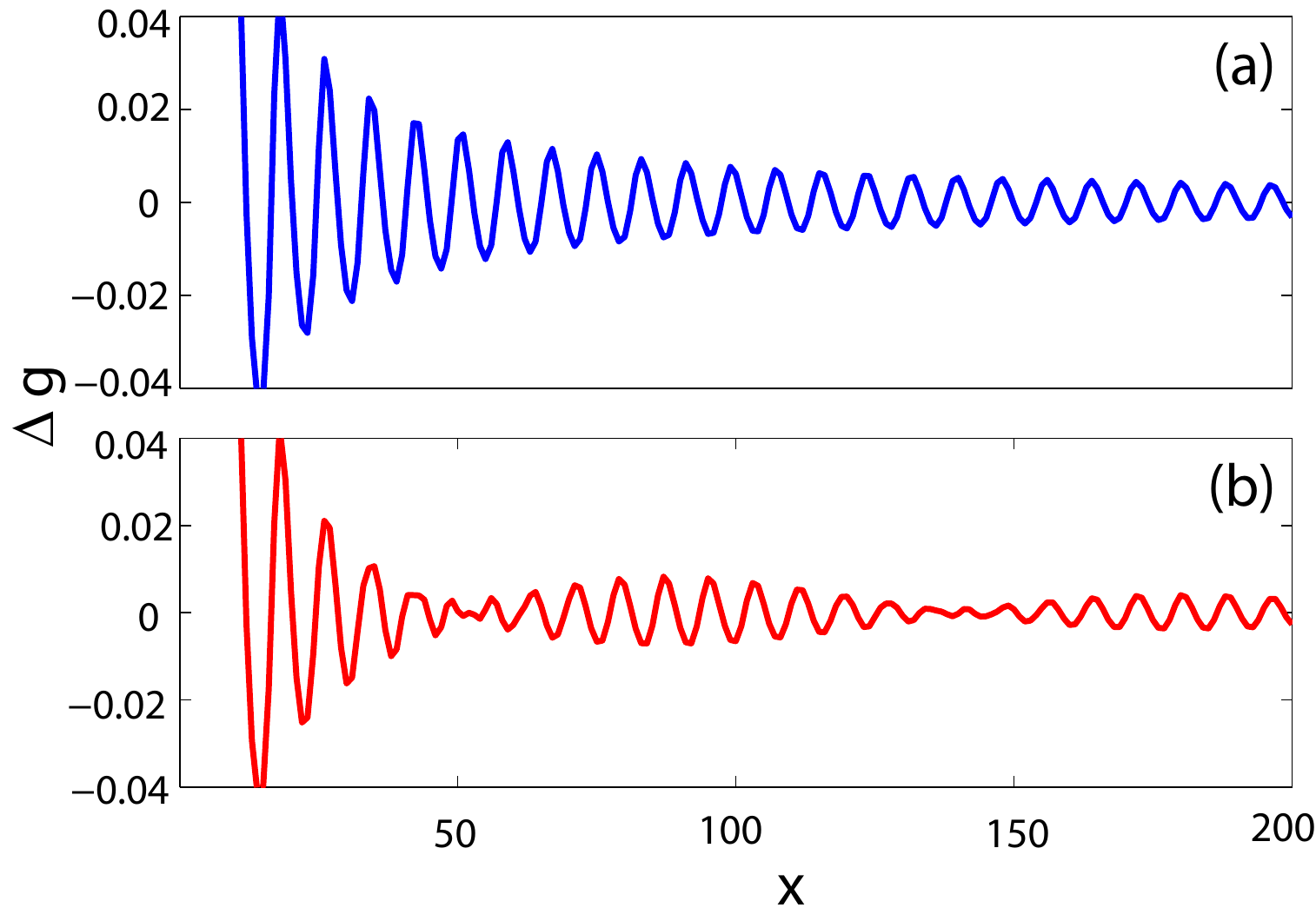}
\caption{(Color online) At ${\cal T}=0$, difference between the effect of a local (a) or global (b) field $h=0.0136$: 
$\Delta g (x,y=0)$ is plotted as a function of $x$ for a tip potential $V=-2$, $E_F=0.1542$, $\Gamma^*=0.0035$ and $\lambda_F/2=8$.}
\label{fig10-a}
\end{figure}
\begin{figure}
\includegraphics[keepaspectratio,width=\columnwidth]{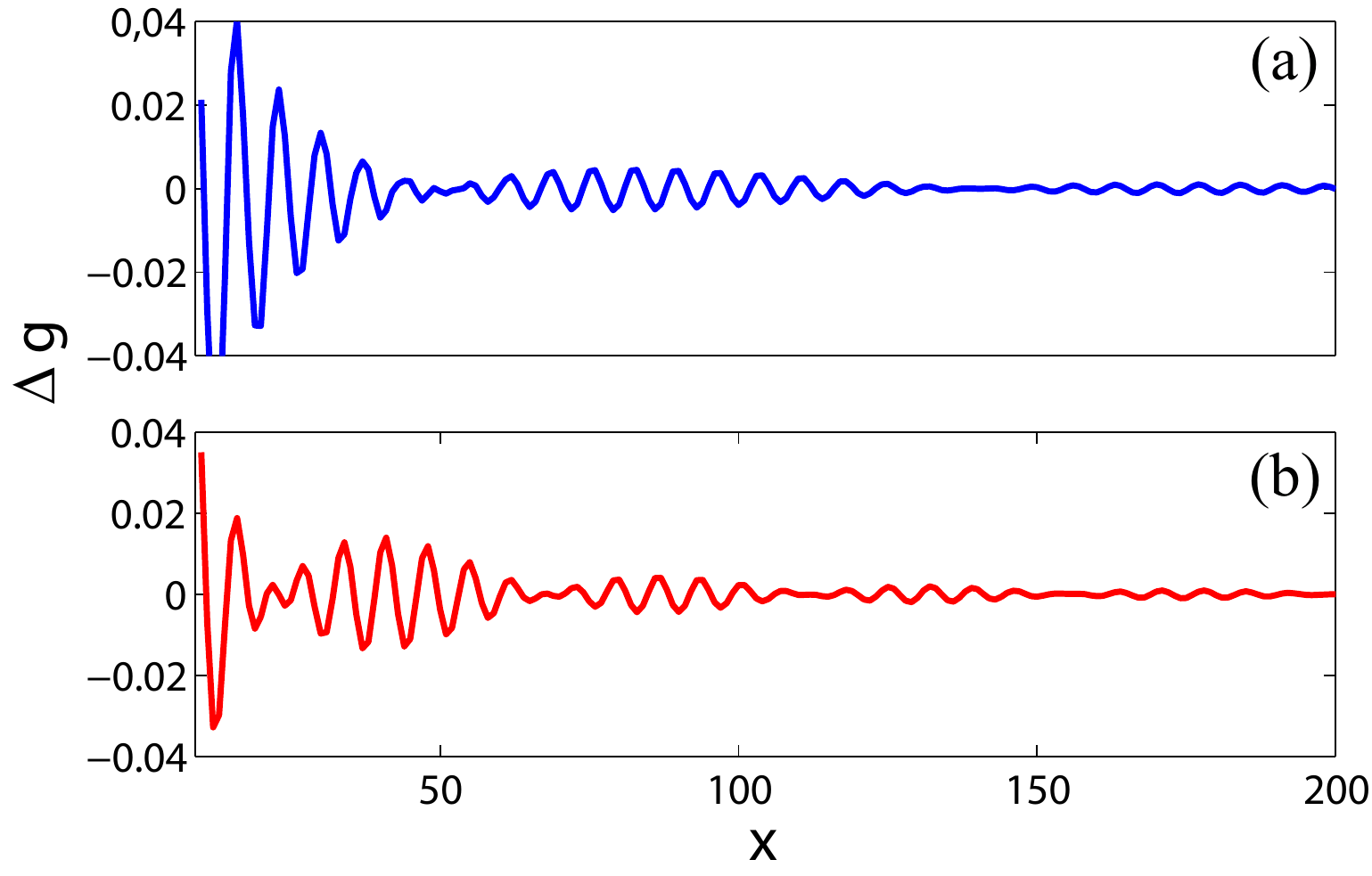}
\caption{(Color online)  RCM contact with parallel magnetic field $h=0.0136$, $E_F=0.1542$, $\Gamma^*=0.0035$, $V=-2$, and $\lambda_F/2=8$: 
At ${\cal T}={\cal T}^*=0.0099/k_B$, $\Delta g (x,y=0)$ is plotted as a function of the tip coordinate $x$ keeping $y=0$. (a): Magnetic field 
inside the contact only ($k_F^{\uparrow}= k_F^{\downarrow}$). (b): Magnetic field everywhere ($k_F^{\uparrow} \neq k_F^{\downarrow}$). 
The difference of periodicity and frequency are mainly due to the field dependence of $k_F^{\sigma}$.} 
\label{fig10-b}
\end{figure}

Let us underline the difference between the effect of a field restricted to the contact or applied everywhere as ${\cal T}\to 0$. 
In the first case (see Eq.~\eqref{radius-ring}), the first ring has a radius  $r^{\cal D}(n=0) \to \infty$ and cannot be seen. 
It is only when one increases ${\cal T} \to {\cal T}^*$ that $r^{\cal D}(n=0)$ becomes small enough and that the rings can be seen. 
In the second case, the rings do not depend on ${\cal T}$ when $r \ll l_{\cal T}$ (see Eq.~\eqref{deltagwithT}) 
and remains visible as  ${\cal T}\to 0$. Numerical illustrations of the differences between the SGM images obtained when a field 
is either applied everywhere or restricted to the contact are given in Fig.~\ref{fig10-a} when ${\cal T} = 0$ and in Fig.~\ref{fig10-b} 
when ${\cal T}={\cal T}^*$: In Fig.~\ref{fig10-a}, one can notice the absence of a beating effect when the field is applied only inside 
the contact, while it can be seen if the field is applied everywhere. In  Fig.~\ref{fig10-b}, the beating effect is visible in the two 
cases.


\section{SGM of a double-dot setup}
 
\begin{figure}
\includegraphics[keepaspectratio,width=\columnwidth]{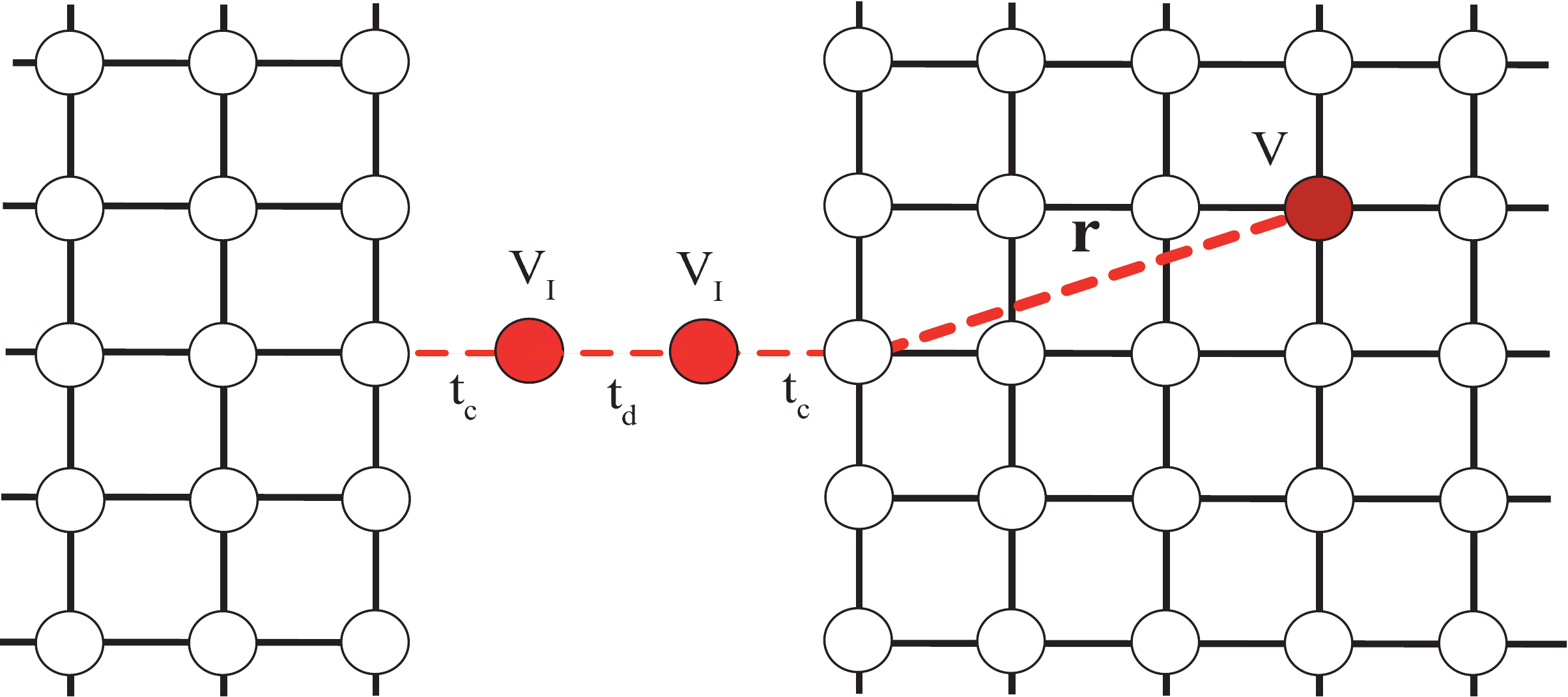}
\caption{(Color online) Scheme of the SGM of a contact made of a double-dot setup.} 
\label{fig12_1}
\end{figure}

 Instead of using a magnetic field for breaking the spin degeneracy of a single resonance, let us now show that  
a contact having a double-peak structure of its transmission without magnetic field gives rise also to a similar 
beating pattern when it is opened between the peaks. Let us take a contact made of two sites of potentials $V_{I}-4t$ 
coupled by an hopping term $t_d$ (see Fig.~\ref{fig12_1}). This gives rise to a two-level system which is often 
used~\cite{revmodphys} to describe electron transport through double quantum dots. As before, the depletion region 
induced by the charged tip is described by a single scattering site and we take again $t=-1$. 
\begin{figure}
\includegraphics[keepaspectratio,width=\columnwidth]{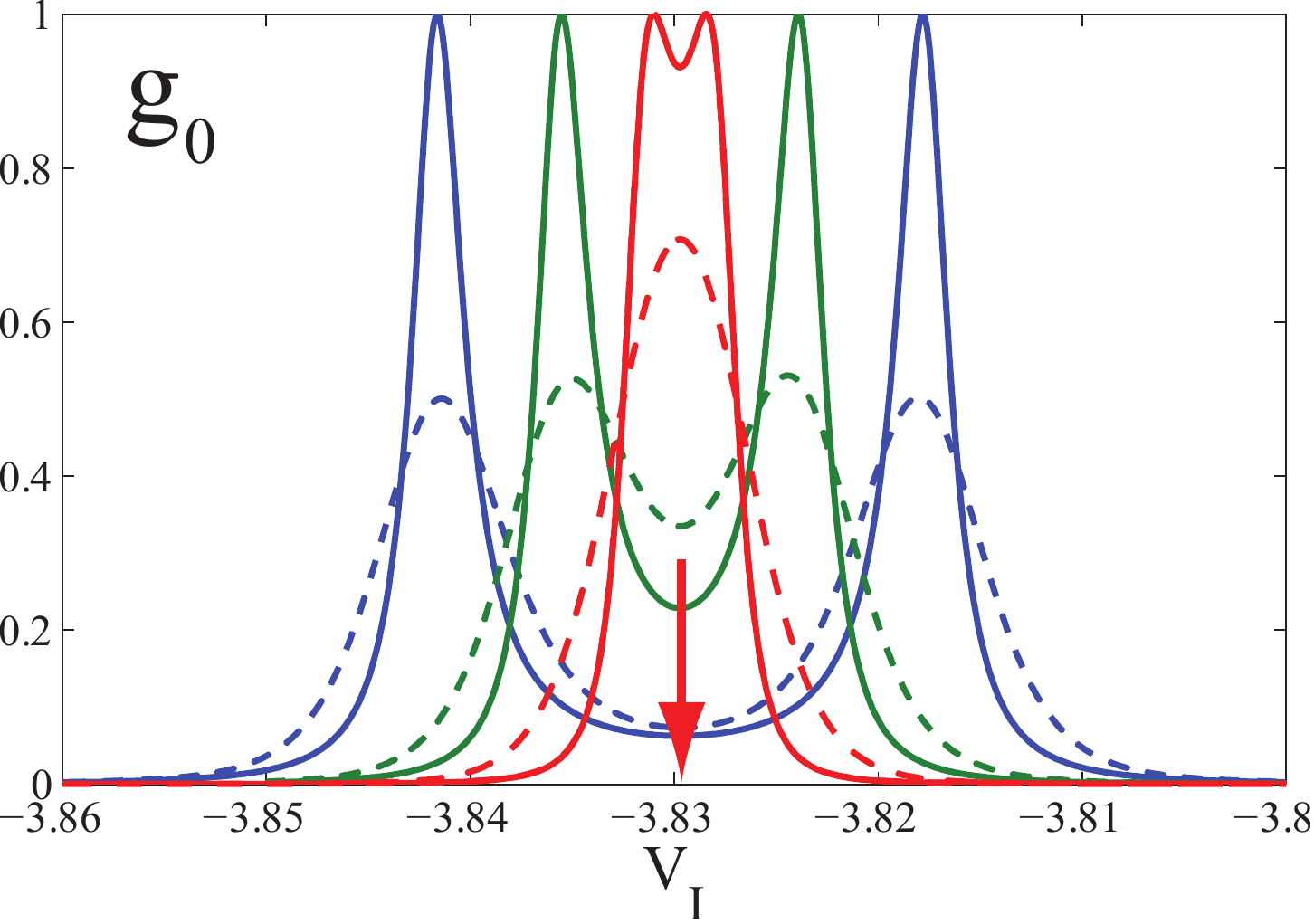}
\caption{(Color online) Double-dot setup without tip ($V=0$): Conductance $g_0$ (in units of $2e^2/h$) with $t_c=0.2$ and $\Gamma=0.0015$ 
as a function of the dot potentials $V_{\bf I}$ for $t_d=0.012$ (blue) $t_d=0.006$ (green) and $t_d=0.002$ (red). 
${\cal T}=0$ (solid line) and $k_B{\cal T}=\Gamma/2$ (dashed line). The arrow gives the potential $V_{\bf I}$ for the SGM study.} 
\label{fig12_2}
\end{figure}
\begin{figure}
\includegraphics[keepaspectratio,width=\columnwidth]{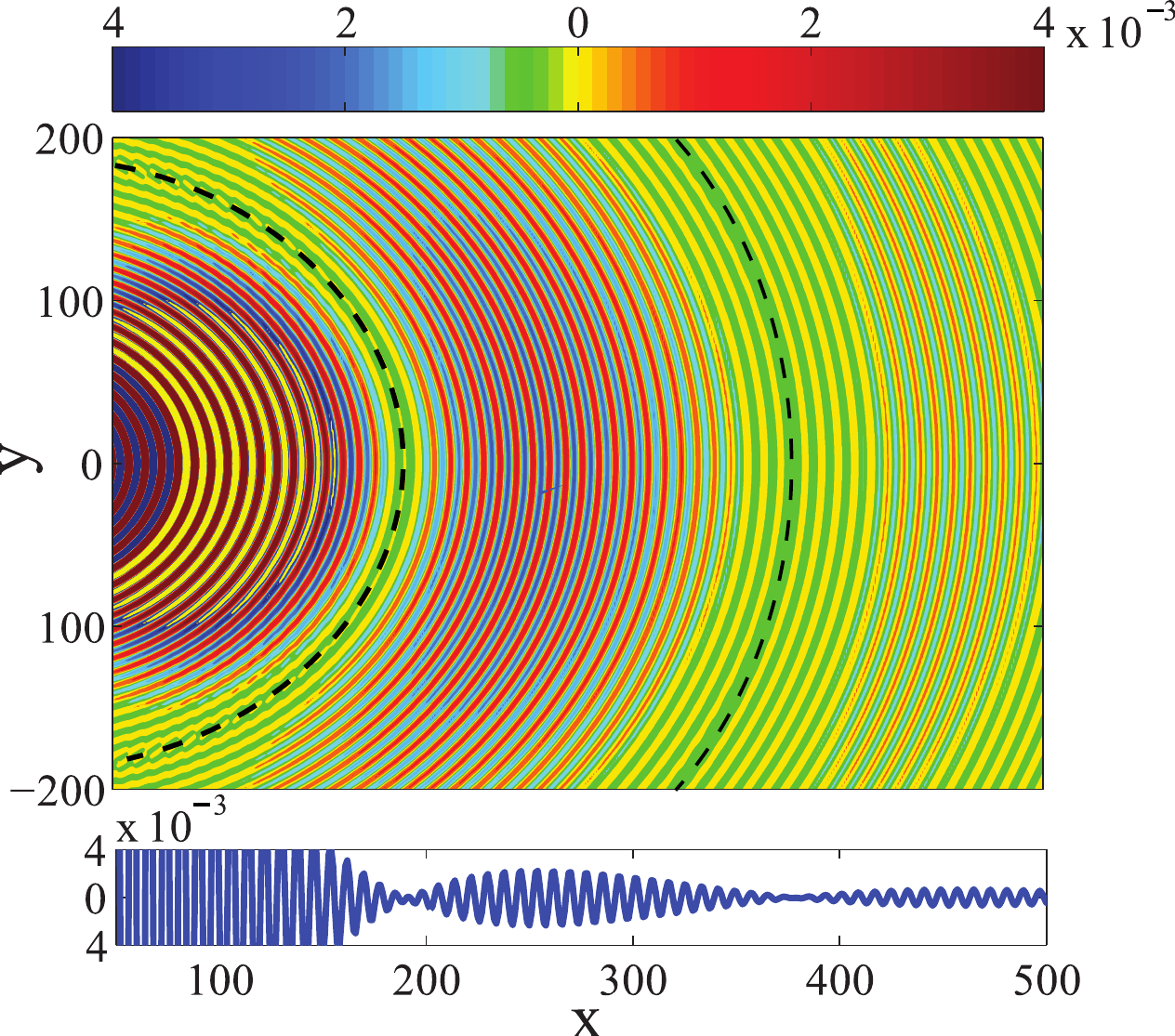}
\caption{(Color online) Double-dot setup - Main Fig: $\Delta g/g_0$ as a function of the tip coordinates $(x,y)$ at a temperature 
${\cal T}={\cal T}^*=0.0022/k_B$ where $l_{\cal T}=78$ ($g_0=0.3064$, $\Gamma=0.0015$, $V=-2$, $t_d=0.006$, $t_c=0.2$, $E_F=0.1542$ and $\lambda_F/2=8$). 
The dashed lines give the circles of radii $r^{\cal D}(n)$ predicted by the theory (Eq.~\eqref{radius-ring} after making the changes 
$h \to t_d$ and $\Gamma \to \Gamma/2$). Above: Color code giving the magnitude of the relative effect. 
Below: $\Delta g/g_0(y=0)$ as a function of $x$ (same parameters as in the main figure).} 
\label{fig13}
\end{figure}
\begin{figure}
\includegraphics[keepaspectratio,width=\columnwidth]{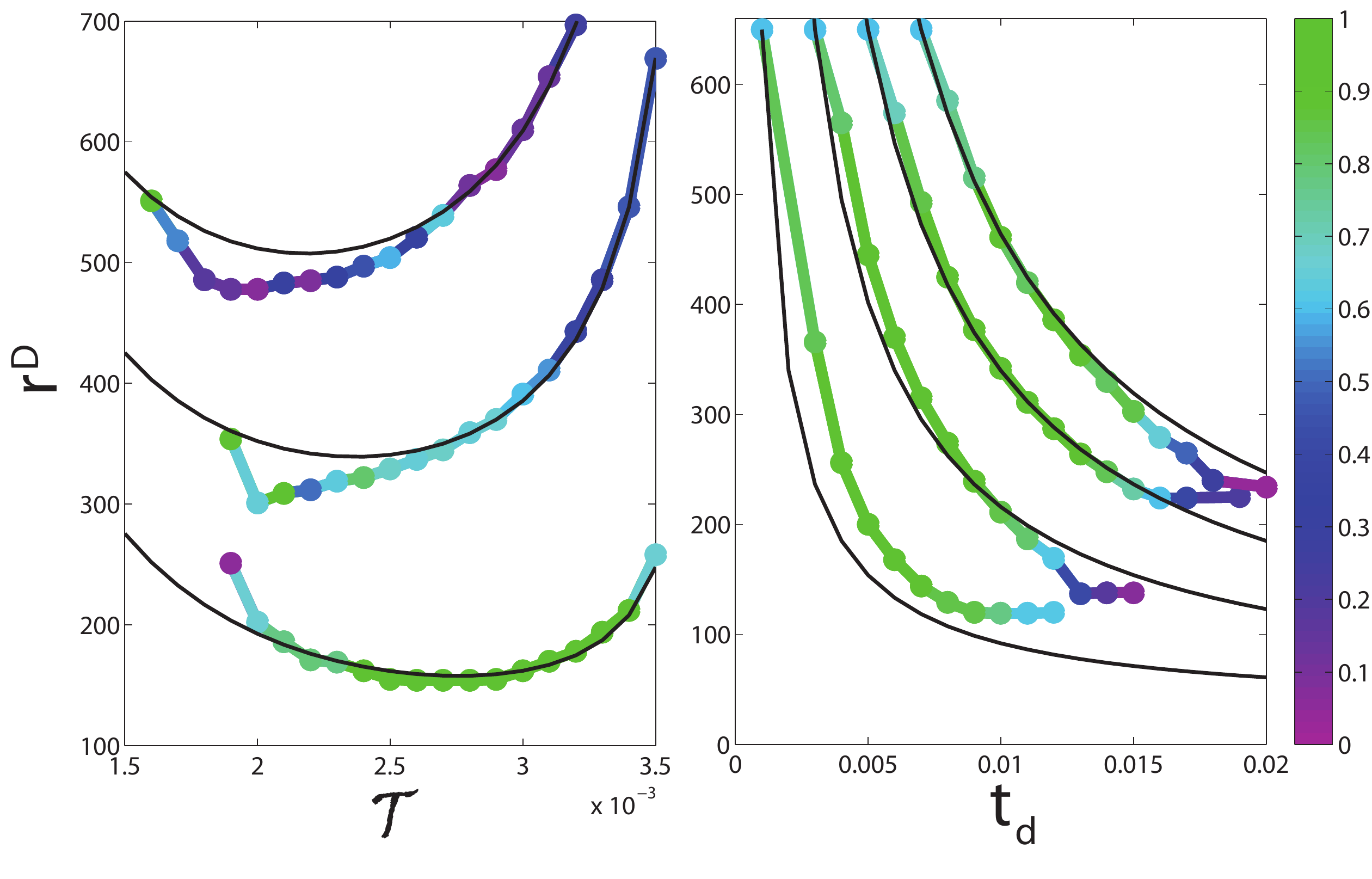}
\caption{(Color online) Double-dot setup with $\Gamma=0.003$: Radii $r^{\cal D}(n)$ of the interference rings for $n=0, 1, 2$ and $3$ 
as a function of $k_B{\cal T}$ (left fig. $t_d=0.01$) and of the interdot coupling $t_d$ (right fig. $k_B{\cal T}=0.0028$).  The dots 
give the successive values of the radii where the numerically calculated values of $\Delta g/g_0 \approx 0$, the colors corresponding 
to a visibility scale indicated at the right. The solid lines are the analytical values of $r^{\cal D}(n)$ derived assuming $r^{\cal D}(n)>r^*$.} 
\label{fig14}
\end{figure}
The Green's function of this model is now given by a $2 \times 2$ matrix which reads 
\begin{equation*} 
G(E)=  
\begin{bmatrix}
E-4-V_{I}-\Sigma_l(E) & -t_d \\
 -t_d & E-4-V_{I}-\Sigma_r(E)-\Delta\Sigma_r \\
\end{bmatrix}
\end{equation*}
where $\Sigma_{r,l}=R+iI$ are the lead self-energies and $\Delta\Sigma_r$ the change induced on $\Sigma_r$ by the tip. 
Without tip, the transmission of an electron of spin $\sigma$ reads
\begin{equation}
T_0^{\sigma}(E)= \frac{t_d}{{\tilde E}}\left(\frac{I^2}{(t_d-{\tilde E})^2+I^2}-\frac{I^2}{(t_d-{\tilde E})^2+I^2}\right)
\label{2QT-transmission_without_tip}
\end{equation}
where ${\tilde E}=E-4-V_{\bf I}-R$.  
This gives two peaks of equal height and width $\Gamma/2=-I$ (instead of $\Gamma$ for the RCM model) 
which are spaced by a ``Zeeman energy'' $2 t_d$ (instead of $2h$ for the RCM model with a field restricted to the contact). 
Of course, the transmission of the double-dot setup is suppressed when the inter-dot coupling $t_d \to 0$, in contrast to the 
RCM model through which the electrons are transmitted when $h \to 0$.\\
\indent These similarities are a consequence of the inversion symmetry of the double-dot model. If the two sites of the contact have 
respective coordinates $(0,0)$ and $(1,0)$, one can rewrite the Hamiltonian of the double-dot setup in terms of fermion operators
which destroy/create an electron of spin $\sigma$ in an even/odd combination of two symmetric orbitals of the original model. 
For instance, for the destruction operators of a particle with spin $\sigma$ and pseudospin $e,o$, one has
\begin{eqnarray}
a^{\sigma,o}_{(x,y)}=(c^{\sigma}_{(-x+1,-y)}-c^{\sigma}_{(x,y)})/\sqrt{2}\\
a^{\sigma,e}_{(x,y)}=(c^{\sigma}_{(-x+1,-y)}+c^{\sigma}_{(x,y)})/\sqrt{2}.
\end{eqnarray}
This allows us to map the original model (electrons with spins free to move on two semi-infinite square lattices coupled by 
two sites of potential $V_{I}$) onto a transformed model of electrons with spins and pseudo-spins (even or odd states) 
free to move on a single semi-infinite square lattice coupled by a single hopping term $t_c$ to a single site of potential 
$V_{I} \pm t_d$ ($+ t_d$ for the odd states, $-t_d$ for the even states). In that sense, $t_d$ can be seen  
indeed as a ``pseudo Zeeman energy'' which removes the pseudo-spin degeneracy. For more details about this mapping, we 
refer the reader to Ref.~\cite{fp1}, where an inversion symmetric one dimensional model was studied, the extension to two dimensions 
being straightforward.\\
\indent The similarity between the two models is also evident when one compares the conductance $g_0$ (in units of $2e^2/h$) 
of the double-dot setup shown in Fig.~\ref{fig12_2} for different values of $t_d$ and ${\cal T}$ with the conductance 
$g_0$ (in units of $e^2/h$) of the RCM model for different values of $h$ and ${\cal T}$. This leads us to expect that their SGM 
images must be also similar: When $t_d \neq 0$, the SGM images of the double-dot setup should also exhibit rings where the effect 
of the tip does not change the conductance of the contact, the radii of these rings being given by Eq.~\eqref{radius-ring}, 
after making the changes $h \to t_d$ and $\Gamma \to \Gamma/2$. This replacement implies also that the ring spacing is equal to 
$\pi k_F/t_d$ for the double-dot setup. A numerical check of such a prediction is given in Fig.~\ref{fig13} (which shows that the 
SGM of the double-dot setup gives rise to a similar pattern of rings) and in Fig.~\ref{fig14} (which shows that the location of 
the rings is indeed given by Eq.~\eqref{radius-ring} when one puts $t_d$ instead of $h$ and $\Gamma/2$ instead of $\Gamma$).


\section{SGM of a QPC with quantized conductance plateaus}
\label{QPC} 
\begin{figure}
\includegraphics[keepaspectratio,width=\columnwidth]{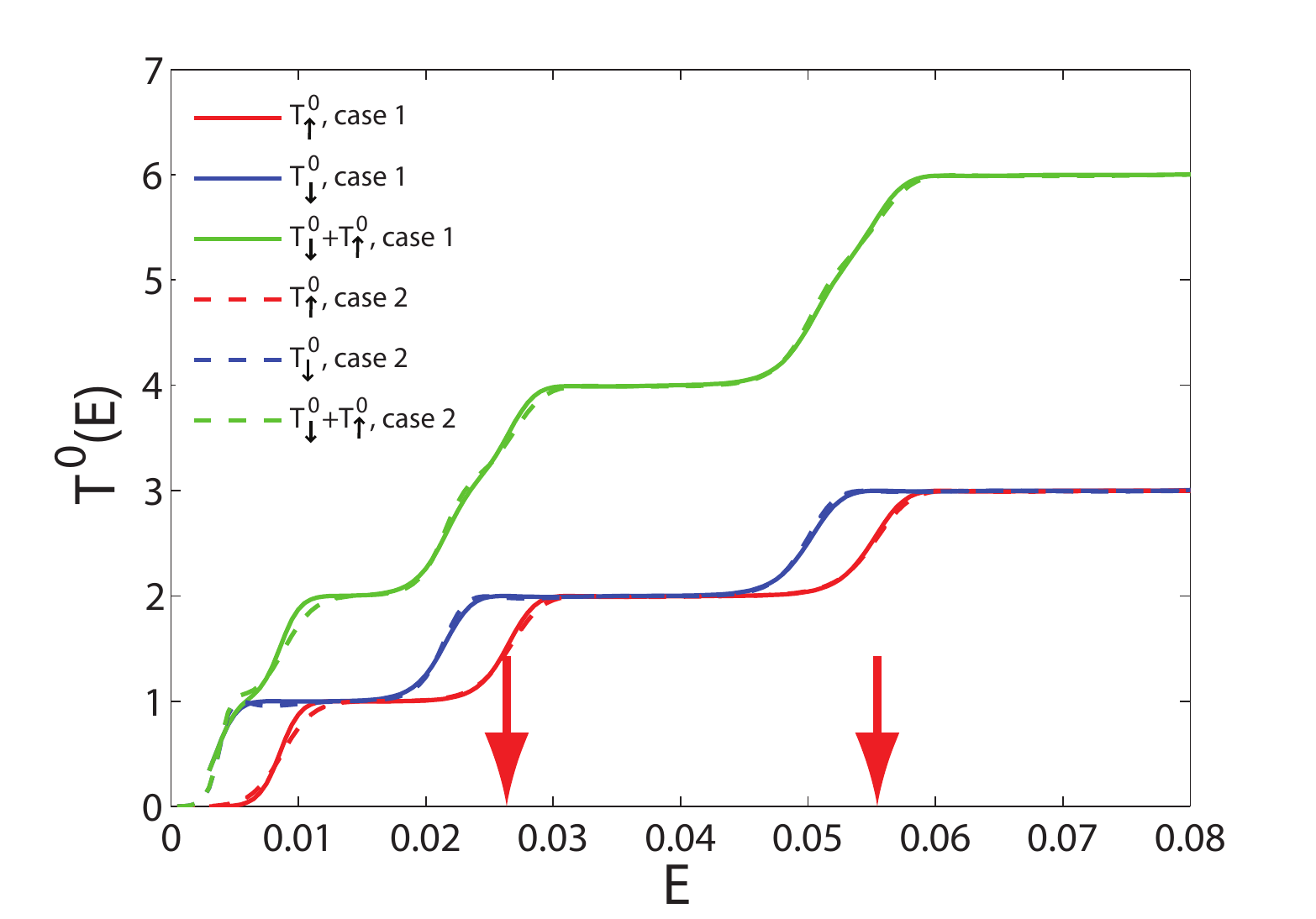}
\caption{(Color online) Lattice model (see Fig.~\ref{fig2QPC}) of a QPC without tip ($V=0$) with $L_y=20$, $L_x=16$ and $k=2$: 
Transmission $T_{\sigma}^0$ of an electron of spin $\sigma$ and total transmission $T^{0}=\sum_{\sigma}T_{\sigma}^0$ as a function 
of $E$ when a parallel magnetic field $h=0.0025$ is applied  everywhere (case 1) and only upon the contact region $-L_x \leq x \leq L_x$ 
(case 2). The arrows give energies around which beating effects can be seen.} 
\label{fig16}
\end{figure}
\begin{figure*}
\includegraphics[keepaspectratio,width=\textwidth]{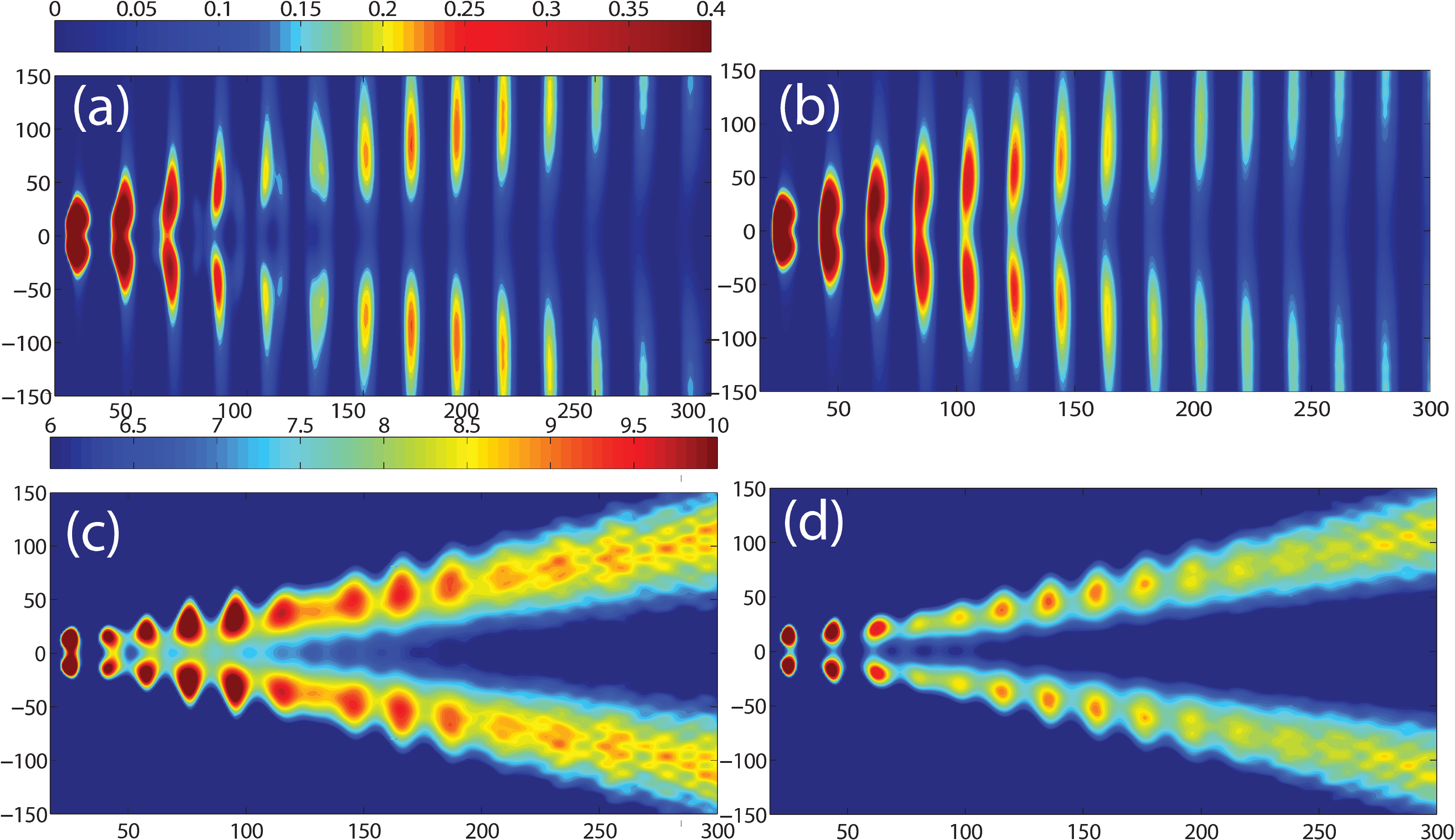}
\caption{(Color online). SGM images of a QPC at the opening of the second transmission channel  ($E_F=0.027$, left arrow of Fig.~\ref{fig16}) 
at ${\cal T}=0$ (top) and $k_B{\cal T}=0.006$ (bottom). A parallel magnetic field $h=0.0025$ is applied everywhere (left) or only upon the contact 
region (right). For a better visibility, we have plotted as a function of the tip coordinates $x,y$, $\frac{\delta (\Delta g)}{\delta x} \times r$ 
when ${\cal T}=0$ and $\frac{\delta (\Delta g)}{\delta x} \times r^2$ when $k_B{\cal T}=0.006$.} 
\label{fig17}
\end{figure*}

We now consider larger contacts, able to have more than one open transmission channel. The energy dependence of their transmission 
is given by a staircase function, in contrast to the single Breit-Wigner resonance of the RCM model without field. Such staircase 
functions with quantized conductance plateaus are observed in the setups sketched in Fig.~\ref{fig1} where the gate potentials give 
rise to a smooth saddle-point QPC potential. These staircase functions characterize also the lattice model for a QPC sketched in 
fig.~\ref{fig2QPC}, as one can see in Fig.~\ref{fig16} in the cases where a parallel magnetic field is applied either everywhere, 
or only inside the contact region.  

An analytical approach being more difficult, we numerically study these larger contacts. The self-energies 
$\Sigma_{l,r}(E)$ are now two matrices of size $(2 L_y+1) \times (2 L_y+1)$, where $L_y$ defines the width of 
the contact region (see Fig.~\ref{fig2QPC}). Their analytical expressions without tip can be found in 
Ref.~\cite{Datta:book97}. To include the effect of the tip which modifies the potential of a single site in the right 
lead, we use again Dyson equation ($V$ playing the role of a perturbation), extending the method used for the RCM model. 
The usual recursive numerical method for calculating the Green function is only used for the contact region of width 
$\leq 2 L_y+1$. With this method, we study leads of very large transverse width ${\cal L}_y \approx 2. 10^4$ 
and a contact region of size $L_x=16$ and $L_y=20$  where the potential $V_{\bf i}$ of a contact site ${\bf i}$ of coordinates 
$(i_x,i_y)$ is taken infinite if $|i_y| \geq (L_y-k)+k \left(i_x/L_x \right)^2$. Once the self energies $\Sigma_{l,r}(E)$ 
of the leads are obtained, the total interferometer transmission $T(E)$ is calculated.

The channel openings of these QPCs play the role of the resonances of the RCM contact~\cite{alp} and one can also use a 
parallel magnetic field to split the QPC channel openings. The effect of a parallel magnetic field upon the SGM images 
of a QPC is shown in Fig.~\ref{fig16}, both when it is applied everywhere or restricted to the contact region. For a QPC 
opened in the energy interval where a new channel is opened for the electrons with parallel spin, but not yet for those 
with antiparallel spins, the effect of the tip upon the QPC conductance should exhibit a beating between the two spin 
contributions of this new channels. This is indeed what can be seen in Fig.~\ref{fig17} when the contact is 
open around the second channel opening (see the arrow in Fig.~\ref{fig16}). The effect of the tip has the V-shape 
which characterizes this second channel. To make the SGM images clearer, we have plotted the effect of the tip over 
the conductance derivatives with respect to $x$. In panels (a) and (b) of Fig.~\ref{fig17}, the SGM images are taken at 
${\cal T}=0$. When the field is applied everywhere (a), one can see a beating effect between the two spin contributions of 
the second channel. When the field is restricted to the contact region (b), there is no beating effect at ${\cal T}=0$. 
In panels (c) and (d), the SGM images are taken at a temperature where $k_B{\cal T}=0.006$, and one recovers a beating effect 
when the field is applied only in the contact region: The behaviors which were previously explained using the RCM contact 
can be qualitatively extended to a QPC with quantized conduction modes. We observed similar beating patterns near the 
opening of the other channels.
  

\section{Conclusion}
 
In summary, we have studied contacts open around spin-degenerate resonances or channel openings. 
Breaking the spin degeneracy by a magnetic field $h$ (or the pseudo-spin degeneracy by an inter-dot 
coupling $t_d$), we have shown that the SGM images of contacts open between split resonances 
or channel openings exhibit a beating pattern of interference fringes where the effect of the tip 
is suppressed. The spacing of the Fabry-P\'erot fringes being $\lambda_F/2$, SGM provides a method for 
measuring by electron interferometry both $\lambda_F$ and either the magnetic field $h$ (RCM-contact) 
or the inter-dot coupling $t_d$ (double-dot setup). We have shown that the spacing between the rings 
is given respectively by $\pi k_F/h$ and $\pi k_F/t_d$.\\
\indent If the magnetic field is applied everywhere, these rings can be observed at zero temperature. If the field is 
applied only upon the contact, the rings are sufficiently close to the contact for being observable only if the 
temperature is close to an optimum temperature ${\cal T}^*$ where $k_B{\cal T}^*$ is of order of the peak splitting.\\
\indent These beating effects can be seen in the SGM images of the RCM model (with a single resonance) or of a QPC (with 
quantized conductance plateaus), when the spin degeneracy is removed by a field. The difference between a local or 
a global Zeeman effect which was analytically shown using the RCM model can be also seen in the numerical sudies of QPCs.\\
\acknowledgments{This research has been supported by the EU Marie Curie network ``NanoCTM'' (project no.234970). 
Discussions with B. Brun, K. Ensslin, M. Sanquer and H. Sellier about SGM experiments and 
with C. Gorini, R. Jalabert and D. Weinmann about SGM theory are gratefully acknowledged.} 

\bibliography{PRB_KLFP}
\end{document}